\documentclass{emulateapj}
\usepackage{amsmath}

\usepackage{color}

\def\etal{{et~al.}}
\def\kms{{\hbox{km s$^{-1}$}}}
\def\ewha{{\hbox{EW(H$\alpha$)}}}

\shorttitle{The IRX-$\beta$ Relation for Local Galaxies}
\shortauthors{Grasha \etal}

\begin{document}
\title{The Nature of the Second Parameter in the IRX-$\beta$ Relation for Local Galaxies}
\author{Kathryn Grasha\altaffilmark{1}, Daniela Calzetti\altaffilmark{1}, Jennifer E. Andrews\altaffilmark{1}, Janice C. Lee\altaffilmark{2}, Daniel A. Dale\altaffilmark{3}} 
\altaffiltext{1}{Astronomy Department, University of Massachusetts, Amherst, MA 01003, USA; kgrasha@astro.umass.edu}
\altaffiltext{2}{STScI, 3700 San Martin Drive, Baltimore, MD 21218, USA}
\altaffiltext{3}{Department of Physics, University of Wyoming, Laramie, WY 82071, USA}

\begin{abstract}
We present an analysis of 98 galaxies of low-dust content, selected from the $Spitzer$ Local Volume Legacy survey, aimed at examining the relation between the ultraviolet (UV) color and dust attenuation in normal star-forming galaxies.  The IRX-$\beta$ diagram relates the total dust attenuation in a galaxy, traced by the far-IR (FIR) to UV ratio, to the observed UV color, indicated by $\beta$.  Previous research has indicated that while starburst galaxies exhibit a relatively tight IRX-$\beta$ relation, normal star-forming galaxies do not, and have a much larger spread in the total-IR to far-UV (FUV) luminosity for a fixed UV color.  We examine the role that the age of the stellar population plays as the ``second parameter' responsible for the observed deviation and spread of star-forming galaxies from the starburst relation.  We model the FUV to FIR spectral energy distribution (SED) of each galaxy according to two broad bins of star formation history (SFH):  constant and instantaneous burst.  We find clear trends between stellar population mean age estimators (extinction-corrected FUV/NIR, U$-$B, and \ewha) and the UV color $\beta$; the trends are mostly driven by the galaxies best-described by instantaneous burst populations.  We also find a significant correlation between $\beta$ and the mean age directly determined from the best-fit instantaneous models.  As already indicated by other authors, the UV attenuation in star-forming galaxies may not be recovered with the UV color alone and is highly influenced by the stellar population's mean age and SFH.  Overall, the scatter in the IRX-$\beta$ diagram is better correlated with $\beta$ than with the perpendicular distance, $d_p$.
\end{abstract}

\section{Introduction}
Obtaining detailed knowledge of the stellar populations of galaxies is one way to gain insight into the evolution and formation of galaxies in the universe.  Studying the stellar populations that give rise to the observed spectral energy distribution (SED) of a galaxy supplies estimates of the star formation rate (SFR) along with the star formation history (SFH).  A major obstacle to understanding the intrinsic SED of a galaxy is correcting for the ultraviolet (UV) and optical flux lost to dust attenuation, which is re-emitted in the infrared (IR).  UV wavelengths, while the most susceptible to the effects of dust, constrain properties of the young stellar populations; longer wavelengths provide information on the older and more evolved stellar populations.  Observations from the UV to the far-IR (FIR) can be used to constrain the amount of dust attenuation in a galaxy.

At high-redshift, where multi-wavelength information is often limited, correcting for dust attenuation in galaxies is commonly done with the so-called IRX-$\beta$ relation, which relates the observed UV spectral slope ($\beta$), or the UV color, to the fraction of UV stellar emission absorbed by dust and re-emitted in the IR (expressed as the ratio $L_{\rm TIR}/L_{\rm FUV}$, where $L_{\rm TIR}$ is the total dust luminosity in the IR and $L_{\rm FUV}$ is the observed stellar emission at $\lambda \approx 1500$~\AA).  These two quantities are correlated in local starburst galaxies \citep{meurer99} and in high-redshift systems \citep{reddy10,reddy12}.  Therefore, the measurement of the UV spectral slope or color immediately yields an estimate of the total dust attenuation in the UV \citep{calzetti00}.  Conversely, no such tight correlation is found in more quiescent, ``normal'' star-forming galaxies, characterized by a SFR/area of $\sum_{\rm SFR}<0.1$ M$_{\sun}$~yr$^{-1}$ kpc$^{-2}$ \citep[e.g,][]{calzetti05} and global SFRs of a few M$_{\sun}$~yr$^{-1}$ or less.  Additionally, their SEDs are not dominated by a central AGN.  We term these systems ``normal star-forming galaxies'' from now on.  

Since the deviations of normal star-forming galaxies from the IRX-$\beta$ relation was discovered \citep{buat02}, great effort has been dedicated to studying the IRX-$\beta$ diagram using the global flux of galaxies \citep{kong04,seibert05,buat05,johnson07,dale07,dale09,gildepaz07,hao11} and spatially resolved galactic regions \citep{bell02,gordon04,calzetti05,popescu05,perez06,thilker07,munoz09,boquien09,boquien12,mao12}.  The normal star-forming galaxies on the IRX-$\beta$ diagram show an order of magnitude scatter compared to the relatively tight starburst IRX-$\beta$ relation.  Currently, no accurate, small-scatter IRX-$\beta$ relation for normal star-forming galaxies exists nor is there agreement on the underlying physical cause of the spread.  

The main difficulty of the studies of normal star-forming galaxies conducted so far is discriminating effects of dust attenuation (variations in dust geometry and extinction curve) from effects of age spread/SFH in the stellar populations.  The main reason is that changing dust geometry and a larger contribution from evolved stellar populations can both make $\beta$ larger.  Extant studies tend to adopt simplifying assumptions for the dust geometry, typically a foreground distribution, which is then parametrized to allow variations in the effective attenuation \citep[e.g.,][]{boquien12}.  These assumptions can have effects on the final results that can be difficult to test or control, as complex dust geometries affect $\beta$ in ways that foreground geometries cannot implement \citep[e.g.,][]{calzetti01,calzetti05}.  

\citet{kong04} were the first to suggest that the deviations of normal star-forming galaxies from the IRX-$\beta$ relation of starbursts could be due to the presence of a ``second parameter''.  They suggested this second parameter to be the birthrate parameter, $b$, which is the ratio of present-to-past SFR.  In this scenario, starbursts, which have high-$b$ values, follow a specific relation in the IRX-$\beta$ plane, driven simply by an increase in total dust obscuration for redder UV colors (larger UV slope $\beta$).  As the birthrate values decrease, galaxies progressively deviate from the starburst IRX-$\beta$ relation in a direction, $d_p$, perpendicular to the relation itself.  That is, at a given total attenuation $L_{\rm IR}/L_{\rm UV}$, a galaxy with a lower $b$ would be intrinsically redder in the UV than one with a high-$b$, due to the contribution of older stellar populations to the UV flux.  In a similar venue, \citet{cortese08} and \citet{dale07,dale09} suggested that the mean age of the stellar population, as traced by the near UV-to-near IR stellar flux ratio and/or by the H$\alpha$ equivalent width, \ewha, is the second parameter.  In a recent study of resolved regions in nearby galaxies, \citet{boquien12} find that the deviations from the IRX-$\beta$ relation are driven mainly by the SFH of the region, which determines the intrinsic value of the UV slope (larger for more evolved regions).  They also find that variations in the extinction are a secondary effect, and that ``traditional'' tracers of the population's mean age, e.g., D$_n$(4000) \citep[the spectral break around 4000~\AA, e.g.,][]{kauffmann03}, and the birthrate parameter are poor predictors of the intrinsic UV slope.  \citet{seibert05,cortese06}, and \citet{johnson07}, indeed, do not find a direct correlation between mean stellar population age indicators and the spread in the IRX-$\beta$ relation.  Interestingly, \citet{reddy10,reddy12} find that at high-redshift, the galaxies that deviate from the IRX-$\beta$ relation of starbursts are those characterized by young ($\lesssim$100~Myr) stellar populations, rather than by more evolved populations, as typical in the local Universe.  All in all, $\beta$ is a poor predictor of the amount of dust attenuation in normal star-forming galaxies \citep{hao11}. 

A number of authors, however, find that the spread in the IRX-$\beta$ relation can be attributed to either variations in the dust extinction properties \citep{gordon00,burgarella05,panuzzo07,inoue06,gildepaz07,boquien09,boquien12} or to both age and extinction variations \citep{mao12}.  \citet{burgarella05,johnson07} and \citet{boquien12} find that larger $\beta$ values correspond to steeper attenuation, suggesting that variations in the adopted attenuation curve could partly be responsible for the observed spread in galaxies, especially in the locus of the IRX-$\beta$ diagram corresponding to large values of the overall dust attenuation.  \citet{calzetti01,boquien09}, and \citet{reddy10,reddy12} showed that adopting different extinction curves and dust geometries can impact the scatter on the IRX-$\beta$ diagram.

Recently, \citet{overzier11} and \citet{takeuchi12} revisited the derivation of the starburst IRX-$\beta$ relation, in order to assess the impact of aperture mis-match in the original data used by \citet{meurer99}.  These authors combined UV data from the \emph{International Ultraviolet Explorer} ($IUE$; 20 x 10 arcsec$^2$ aperture) with IR photometry from the \emph{Infrared Astronomical Satellite} ($IRAS$; 1.5 x 4.7 arcmin$^2$ at 60~$\micron$).  \citet{takeuchi12} finds that using aperture-matched data does indeed reduce the overall discrepancy between starbursts and normal star-forming galaxies, although their highest angular resolution, 50'' (driven by the AKARI 140~\micron\ PSF), already includes significant contamination from non-starburst populations in the convolved UV GALEX images.  When only the starburst regions within the galaxies are measured, \citet{overzier11} find that the starburst relation is minimally changed relative to that of \citet{meurer99}.  We adopt this more recent re-derivation by \citet[][which they call `M99 inner']{overzier11} as our default starburst IRX-$\beta$ relation to utilize the advantage of the improved relation and the extensive literature using the Meurer \etal's relation.  However, we recognize that even the Overzier IRX-$\beta$ relation may still suffer from some limitations due to the mis-matched apertures between the UV and IR fluxes. 

Our study aims at investigating and addressing the past difficulties of the spread in the IRX-$\beta$ relation while removing the need to make assumptions on the characteristics of the dust attenuation in the galaxies.  For this goal, we will concentrate on galaxies that span a large range of $\beta$ values, but are virtually dust-free (or contain only a small amount of dust).  Dust-poor star-forming galaxies offer a unique opportunity to explore the impact of stellar population age on the spread in the IRX-$\beta$ relation, by mitigating or eliminating the need for uncertain assumptions on the dust extinction curve and geometry.  This is complementary to all studies attempted so far, which have usually included mostly dusty galaxies (see references above).  We will focus on quantifying the impact of the mean stellar age on the spread in the IRX-$\beta$ relation, but will not attempt to convert the $L_{\rm IR}/L_{\rm UV}$ ratio to an effective UV attenuation, $A_{\rm UV}$.  Our paper is organized as follows:  we present our sample in Section \ref{sec:2}, Section \ref{sec:3} reviews definitions of the IRX-$\beta$ relation, Section \ref{sec:4} describes our modeling and fitting routine, Section \ref{sec:5} presents the initial results and analysis of our study, Section \ref{sec:6} discusses the implications of our immediate work and comparisons to prior studies, and finally, we summarize our conclusions in Section \ref{sec:7}.  

Throughout this paper, we adopt a $\Lambda$CMD concordance cosmology model of $\Omega_{\rm m} = 0.27$, $\Omega_{\Lambda} =0.73$, and $H_\circ = 70$~\kms~Mpc$^{-1}$ \citep{komatsu11}.  All numbers taken from the literature are re-calculated (if necessary) to be consistent with this cosmological model.

\section{Sample Selection and Data}\label{sec:2}
We have selected a subset of virtually dust-free galaxies from the $Spitzer$ Local Volume Legacy \citep[LVL;][]{dale09} survey, which includes predominantly low-metallicity, low-luminosity dwarf and irregular galaxies.  All 258 galaxies in the LVL survey are local ($D<11$~Mpc) galaxies that avoid the Galactic plane ($|b|>20^{\circ}$), and are brighter than $B = 15.5$ magnitude \citep{lee11}.  The LVL sample is built on UV, H$\alpha$, and $HST$ (\emph{Hubble Space Telescope}) imaging from the 11~Mpc H$\alpha$ and Ultraviolet Galactic Survey \citep[11HUGS;][]{kennicutt08,lee11} and the Advanced Camera for Surveys (ACS) Nearby Galactic Survey Treasury \citep[ANGST;][]{dalcanton09}, providing a statistically robust and complete sample of the nearest galaxies to the Milky Way (MW).  The LVL provides an exceptional way to study the star formation activity in a sample of low-mass, low-surface brightness systems that are not flux-limited.  

We select our sample of low-dust content galaxies from LVL by requiring that each galaxy satisfies the constraint $\log L_{\rm TIR}/L_{\rm FUV}\leq0.5$, resulting in a set of 175 galaxies (Figure \ref{fig:1}, top panel).  This threshold is a compromise between a statistically robust number of galaxies, as to not limit our results by small number statistics, and galaxies that can be considered relatively dust-free (at $\log L_{\rm TIR}/L_{\rm FUV}=0.5$, our galaxies are attenuated at most by a factor of 1.8 in the FUV GALEX band).  This criterion removes one degree of freedom in our analysis and enables us to more accurately pin down the role of stellar population ages for non-starburst galaxies on the IRX-$\beta$ relation.

However, low-dust content galaxies are typically low-mass galaxies in the local Universe, and there are no large, massive spirals in our sample.  Our selection criterion thus places us in a complementary portion of the IRX-$\beta$ parameter space for normal star-forming galaxies than studied so far by most authors.  Our low-dust galaxies span a range in SFR, as measured from the dust-corrected FUV, from $3\times10^{-3}$ to 0.01 M$_{\sun}$~yr$^{-1}$, with an average value of SFR$_{\rm FUV} = 0.04$ M$_{\sun}$~yr$^{-1}$ \citep{lee09a}.  Additionally, our galaxies cover a range in SFR/area of $1.3\times10^{-4} < \sum_{\rm SFR} < 8\times10^{-3}$ M$_{\sun}$~yr$^{-1}$ kpc$^{-2}$ \citep{calzetti10} with an average SFR/area of $\sum_{\rm SFR} = 5\times10^{-4}$ M$_{\sun}$~yr$^{-1}$ kpc$^{-2}$, placing our galaxies well in the range of quiescently star-forming systems \citep{calzetti05}.  
 
We require that all the galaxies have available observations at the FUV, NUV, U, B, V, J, H, Ks, IRAC 3.6~\micron, MIPS 24~\micron, 70~\micron, and 160~\micron\ wavelengths.  The central wavelengths of each required pass band are listed in Table \ref{tab:1}.  FUV and NUV data are readily available for all LVL galaxies from GALEX observations \citep{lee11}, while the 3.6 \micron\ to 160 \micron\ photometric observations are from the MIPS/$Spitzer$ \citep[Multiband Imaging Photometry for $Spitzer$;][]{rieke04} and the IRAC/$Spitzer$ \citep[Infrared Array Camera;][]{fazio04}, from Dale et al. (2009).  For the U, B, and V bands we use photometric data from, in order of preference; the Third Reference Catalogue of Bright Galaxies \citep[RC3;][]{devaucouleurs95,corwin94}, the Vatican Advanced Technology Telescope \citep[VATT;][]{taylor05}, and the Sloan Digital Sky Survey \citep[SDSS;][]{abazajian09}.  The SDSS $u',g',r'$ photometry are converted to the Johnson U, B, V magnitude system according to \citet{jester05}.  

U-band observations are required for each galaxy and give the ability to characterize stellar populations over the age range of $0.1-1$~Gyr.  The U$-$B color is also a strong age discriminator \citep{whitmore99}, albeit somewhat less sensitive than the spectroscopic D$_n$(4000) index that directly targets the stellar photospheric break around 4000 \AA\ \citep{kauffmann03}.  All optical photometry are acquired from NED.  J, H, and Ks bands are obtained from the 2 Micron All Sky Survey \citep[2MASS;][]{skutskie06} catalog, where we use the corrected 2MASS-based fluxes calibrated by \citet{dale09}.  We have also collected \ewha\ data for our galaxies (where available) from \citet{kennicutt08}.  

The photometry is corrected for foreground Galactic extinction with a Milky Way extinction curve \citep{schlegel98}.  The extinction correction is directly retrieved from NED for U, B, V, J, H, and Ks.  For the GALEX band-passes, we assume $A_{\lambda}/E(B-V)=3.1$ and derive:  $A_{\rm FUV} = 8.016\ E(B-V)_{\rm MW}$ and $A_{\rm NUV} = 8.087\ E(B-V)_{\rm MW}$, adopting the Milky Way extinction curve parametrized in \citet{fitzpatrick99}.  We do not correct for foreground extinction data beyond the Ks band.  These flux values $F_{\lambda, \rm corr}$ are used in all of our data analysis.  

\begin{deluxetable}{lccc}
\tabletypesize{\scriptsize}
\tablecaption{Multi-Wavelength Data\label{tab:1}} 
\tablecolumns{4}
\tablewidth{0pt}
\tablehead{
\colhead{Band}& 
\colhead{Wavelength} &  
\colhead{Instrument/Survey} &
\colhead{Reference}   
}
\startdata 
FUV & 1520~\AA & GALEX & 1\\
NUV & 2310~\AA & GALEX & 1\\
U & 3660~\AA & RC3/VATT/SDSS & 2, 3, 4\\
B & 4410~\AA & RC3/VATT/SDSS & 2, 3, 4\\
V & 5540~\AA & RC3/VATT/SDSS & 2, 3, 4\\
J & 1.235~\micron & 2MASS & 5\\
H & 1.662~\micron & 2MASS & 5\\
Ks & 2.159~\micron & 2MASS & 5\\
3.6 & 3.6~\micron & IRAC/$Spitzer$ & 5 \\
MIPS 24 & 24~\micron & MIPS/$Spitzer$ & 5 \\
MIPS 70 & 70~\micron & MIPS/$Spitzer$ & 5 \\
MIPS 160 & 160~\micron & MIPS/$Spitzer$ & 5
\enddata
\tablecomments{
Columns list the 
(1) Photometric band, 
(2) Central wavelength of each band, 
(3) Instrument or survey the photometric data came from, and
(4) References for where we acquired our photometric data. \\
References for photometry:
1 -- \citet{lee11}; 
2 -- RC3; \citet{devaucouleurs95,corwin94}; 
3 -- VATT; \citet{taylor05};
4 -- SDSS; \citet{abazajian09}; and 
5 -- \citet{dale09}.  
}
\end{deluxetable}

Our final sample comprises 98 low-dust content galaxies that have the necessary optical band photometry (U, B, and V).  Some of our photometric data have upper limits as a result of non-detections (detection below a 5$\sigma$ threshold) in the IR bands (J band to 160 \micron); we have treated upper limits as appropriate in our analysis.  Table \ref{tab:2} lists all 98 sources used in our analysis of the IRX-$\beta$ diagram. \\

\LongTables
\begin{deluxetable*}{clcccccccc}
\tabletypesize{\scriptsize}
\tablecaption{Galaxy Properties\label{tab:2}}
\tablecolumns{10}
\tablewidth{0pt}
\tablehead{
\colhead{Galaxy}& 
\colhead{R.A.}	& 
\colhead{Dec.}	& 
\colhead{D}		&
\colhead{SFH}	&
\colhead{Age}	&
\colhead{EW(H$\alpha$)}	&  
\colhead{$\log L_{\rm TIR}$}	&
\colhead{IRX}	& 
\colhead{12+log(O/H)}		
\\		
\colhead{}			& 
\colhead{}			&
\colhead{}			&
\colhead{(Mpc)}     & 
\colhead{}			&
\colhead{(Gyr)}		&
\colhead{(\AA)}		&
\colhead{($L_{\sun}$)}	&
\colhead{}			& 
\colhead{}			
}
\startdata 
NGC 24                        	&	00 09 56.5	&	$-$24 57 47	&	8.13	&	B	&	0.5$^{+0.2}_{-0}$	&	16.2	&	9.24(5)	&	0.29(2)		&	8.93(11)$^{1}$\\
NGC 45                        	&	00 14 04.0	&	$-$23 10 55	&	7.07	&	C	&	5		&	31.2	&	8.87(5)	&	$-$0.12(2)	&	\nodata	\\
NGC 55                        	&	00 14 53.6	&	$-$39 11 48	&	2.17	&	B	&	0.7$^{+0}_{-0.2}$	&	32.8	&	7.50(4)	&	0.238(9)	&	8.05(10)$^{2,3}$\\
NGC 59                        	&	00 15 25.1	&	$-$21 26 40	&	5.3		&	B	&	0.5$^{+0.2}_{-0}$	&	35.2	&	8.88(3)	&	0.122(16)	&	8.40(11)$^{4}$\\
IC 1574                       	&	00 43 03.8	&	$-$22 14 49	&	4.92	&	B	&	0.7$^{+0}_{-0.2}$	&	8.65	&	$<$7	&	$<-$0.11	&	\nodata	\\
UGCA 15                       	&	00 49 49.2	&	$-$21 00 54	&	3.34	&	B	&	0.4$^{+0.1}_{-0}$	&	\nodata	&	$<$6.5	&	$<-$0.19	&	\nodata	\\
								&				&				&			&	C	&	5		&			&			&				&\\
NGC 300                       	&	00 54 53.5	&	$-$37 41 04	&	2		&	C	&	5		&	20		&	9.31(5)	&	0.037(2)	&	8.73(4)$^{3,5,6}$\\
UGC 891                       	&	01 21 18.9	&	+12 24 43	&	10.8	&	B	&	0.5$^{+0.5}_{-0}$	&	12.5	&	7.43(6)	&	$-$0.50(3)	&	8.20(10)$^{7,8}$\\
UGC 1104                      	&	01 32 42.5	&	+18 19 02	&	7.5		&	C	&	5		&	20.6	&	7.24(5)	&	$-$0.66(2)	&	7.94(5)$^{9,10}$\\
NGC 598                       	&	01 33 50.9	&	+30 39 37	&	0.84	&	C	&	5		&	20.5	&	8.50(12)&	0.298(2)	&	8.36(5)$^{11,3,12}$\\
NGC 625                       	&	01 35 04.6	&	$-$41 26 10	&	4.07	&	B	&	0.7$^{+0}_{-0.2}$	&	27.9	&	9.01(2)	&	0.416(15)	&	8.10(10)$^{13,4}$\\
UGC 1176                      	&	01 40 09.9	&	+15 54 17	&	9		&	B	&	0.3$^{+0.1}_{-0}$	&	29.6	&	7.49(6)	&	$-$0.45(2)	&	7.97(3)$^{1}$\\
ESO 245$-$G005                  &	01 45 03.7	&	$-$43 35 53	&	4.43	&	C	&	5		&	26.9	&	7.50(4)	&	$-$0.68(2)	&	7.70(10)$^{14,15,4}$\\
NGC 672                       	&	01 47 54.5	&	+27 25 58	&	7.2		&	C	&	5		&	22.8	&	9.40(4)	&	0.204(19)	&	\nodata	\\
ESO 154$-$G023                  &	02 56 50.4	&	$-$54 34 17	&	5.76	&	C	&	5		&	33.6	&	7.97(8)	&	$-$0.362(15)&	7.7(1)$^{1}$\\
NGC 1313                      	&	03 18 16.1	&	$-$66 29 54	&	4.15	&	B	&	0.1$^{+0}_{-0.13}$	&	26.1	&	9.02(3)	&	0.059(9)	&	8.41(9)$^{3,16}$\\
NGC 1311                      	&	03 20 07.0	&	$-$52 11 08	&	5.45	&	C	&	5		&	30.6	&	7.73(4)	&	$-$0.194(19)&	\nodata	\\
NGC 1487                      	&	03 55 46.1	&	$-$42 22 05	&	9.08	&	C	&	5		&	50.9	&	8.34(4)	&	0.107(14)	&	8.2(2)$^{17,18}$\\
NGC 1510                      	&	04 03 32.6	&	$-$43 24 00	&	9.84	&	C	&	5		&	31.3	&	9.37(3)	&	$-$0.0030(15)&	8.4(2)$^{19,17}$\\
NGC 1512                      	&	04 03 54.3	&	$-$43 20 56	&	9.64	&	B	&	0.4$^{+0.3}_{-0}$	&	10.8	&	8.28(4)	&	0.45(2)	&	8.56(12)$^{1}$\\
NGC 1522                      	&	04 06 07.9	&	$-$52 40 06	&	9.32	&	C	&	5		&	55.5	&	8.85(3)	&	0.0761(15)	&	8.20(12)$^{20}$\\
NGC 1705                      	&	04 54 13.5	&	$-$53 21 40	&	5.1		&	C	&	5$^{+0}_{-4}$		&	100		&	7.84(4)	&	$-$0.624(17)&	8.21(5)$^{21,19}$\\
NGC 1744                      	&	04 59 57.8	&	$-$26 01 20	&	7.65	&	B	&	0.2$^{+0.5}_{-0.13}$&	25.6	&	8.64(9)	&	$-$0.29(2)	&	\nodata	\\
								&				&				&			&	C	&	5$^{+0}_{-0}$		&			&			&				&\\
NGC 1800                      	&	05 06 25.7	&	$-$31 57 15	&	8.24	&	B	&	0.5$^{+0.2}_{-0}$	&	26.5	&	8.32(4)	&	0.045(18)	&	8.4(2)$^{19,22}$\\
NGC 2500                      	&	08 01 53.2	&	+50 44 14	&	7.63	&	C	&	5		&	30.7	&	8.70(4)	&	0.173(18)	&	\nodata	\\
NGC 2537                      	&	08 13 14.6	&	+45 59 23	&	6.9		&	C	&	5		&	35.7	&	8.70(4)	&	0.225(17) 	&	7.71(10)$^{23,24}$\\
UGC 4278                      	&	08 13 58.9	&	+45 44 32	&	7.59	&	C	&	5		&	33		&	8.14(4)	&	$-$0.239(13)&	8.08(19)$^{25,26}$\\
NGC 2552                      	&	08 19 20.5	&	+50 00 35	&	7.65	&	C	&	5		&	24.4	&	8.27(7)	&	$-$0.112(14)&	8.39(17)$^{26,27}$\\
UGC 4426                      	&	08 28 28.4	&	+41 51 24	&	10.3	&	B	&	0.4$^{+0.1}_{-0.1}$	&	9.35	&	$<$7.7	&	$<$0.027	&	\nodata	\\
UGC 4459                      	&	08 34 07.2	&	+66 10 54	&	3.56	&	C	&	1		&	122.5	&	6.98(4)	&	$-$0.328(17)&	7.52(8)$^{28,29,27,4}$\\
UGC 4787                       	&	09 07 34.9	&	+33 16 36	&	6.53	&	C	&	5		&	17.4	&	7.42(5)	&	$-$0.19(4)	&	8.2(2)$^{10}$\\
CGCG 035$-$007                  &	09 34 44.7	&   +06 25 32	&	5.17	&	C	&	5		&	15.5	&	6.78(2)	&	$-$0.13(2)	&	\nodata	\\
UGC 5272                      	&	09 50 22.4	&	+31 29 16	&	7.1		&	C	&	0.7$^{+0.3}_{-0}$	&	43.3	&	7.51(4)	&	$-$0.572(18)&	7.83(8)$^{30,31}$\\
UGC 5340                      	&	09 56 45.7	&	+28 49 35	&	5.9		&	C	&	1$^{+0}_{-0.3}$		&	31.4	&	$<$7.1	&	$<-$0.73	&	7.21(3)$^{32,33}$\\
UGC 5336                      	&	09 57 32.0	&	+69 02 45	&	3.7		&	C	&	0.3$^{+0}_{-0.1}$	&	8.65	&	6.49(2)	&	$-$0.97(3)	&	8.7(3)$^{56}$\\
UGC 5364                      	&	09 59 26.5	&	+30 44 47	&	0.69	&	B	&	0.4$^{+0.1}_{-0.12}$&	5.88	&	$<$5.5	&	$-$0.693(4)	&	7.30(5)$^{34,29}$\\
UGC 5373                      	&	10 00 00.1	&	+05 19 56	&	1.44	&	B	&	0.7$^{+0}_{-0.2}$	&	4.81	&	5.89(6)	&	$-$1.28(3)	&	7.84(5)$^{29,35,36,37}$\\
UGC 5423                      	&	10 05 30.6	&	+70 21 52	&	5.3		&	C	&	5		&	25.7	&	6.87(5)	&	$-$0.26(2)	&	7.98(10)$^{38}$\\
UGC 5456                      	&	10 07 19.6	&	+10 21 43	&	3.8		&	C	&	5		&	43.8	&	7.34(3)	&	$-$0.233(15)&	\nodata	\\
NGC 3239                      	&	10 25 04.9	&	+17 09 49	&	8.29	&	C	&	2.5$^{+2.5}_{-1.5}$	&	54.6	&	8.96(4)	&	$-$0.066(16)&	\nodata	\\
NGC 3274                      	&	10 32 17.3	&	+27 40 08	&	6.5		&	C	&	5		&	46.9	&	8.09(4)	&	$-$0.175(13)&	8.3(3)$^{22,27}$\\
UGC 5764                      	&	10 36 43.3	&	+31 32 48	&	7.08	&	C	&	5		&	42.7	&	6.88(6)	&	$-$0.65(3)	&	7.95(4)$^{7,8,33}$\\
UGC 5829                      	&	10 42 41.9	&	+34 26 56	&	7.88	&	C	&	0.7$^{+0.3}_{-0}$	&	35.2	&	8.01(4)	&	$-$0.514(11)&	8.30(10)$^{7,8}$\\
NGC 3344                      	&	10 43 31.2	&	+24 55 20	&	6.64	&	B	&	0.7$^{+0.3}_{-0.2}$	&	27		&	9.43(5)	&	0.38(2)		&	8.76(2)$^{39,40,3}$\\
UGC 5889                      	&	10 47 22.3	&	+14 04 10	&	9.3		&	B	&	0.7$^{+0}_{-0.2}$	&	7.34	&	$<$7.7	&	$-$0.070(4)	&	\nodata	\\
UGC 5923                      	&	10 49 07.6	&	+06 55 02	&	7.16	&	B	&	0.5$^{+0}_{-0.2}$	&	13		&	7.45(4)	&	0.233(13)	&	8.3(2)$^{10}$\\
UGC 5918                      	&	10 49 36.5	&	+65 31 50	&	7.4		&	B	&	0.4$^{+0.1}_{-0.1}$	&	18.1	&	$<$7.3	&	$-$0.064(4)	&	7.84(4)$^{56}$\\
NGC 3432                      	&	10 52 31.1	&	+36 37 08	&	7.89	&	C	&	5		&	54.2	&	9.24(8)	&	0.357(18)	&	\nodata	\\
NGC 3486                      	&	11 00 23.9	&	+28 58 30	&	8.24	&	C	&	5		&	34.7	&	9.37(4)	&	0.22(2)		&	\nodata	\\
UGC 6457                      	&	11 27 12.2	&	$-$00 59 41	&	10.2	&	C	&	5		&	25.2	&	7.36(6)	&	$-$0.44(3)	&	\nodata	\\
UGC 6541                      	&	11 33 28.9	&	+49 14 14	&	3.9		&	C	&	5		&	82.7	&	6.72(4)	&	$-$0.78(5)	&	7.82(6)$^{41,26,42,43}$\\
NGC 3738                      	&	11 35 48.8	&	+54 31 26	&	4.9		&	C	&	5		&	23.9	&	8.16(4)	&	$-$0.029(16)&	8.23(1)$^{44,22,27}$\\
NGC 3741                      	&	11 36 06.2	&	+45 17 01	&	3.19	&	C	&	1$^{+4}_{-0}$		&	54.9	&	6.50(2)	&	$-$0.79(2)	&	8.1(2)$^{45}$\\
UGC 6782                      	&	11 48 57.4	&	+23 50 15	&	14		&	B	&	0.7$^{+0.3}_{-0.2}$	&	\nodata	&	$<$7.8	&	$<$0.011	&	\nodata	\\
UGC 6817                      	&	11 50 53.0	&	+38 52 49	&	2.64	&	B	&	0.3$^{+0.1}_{-0}$	&	25		&	6.37(6)	&	$-$0.83(3)	&	7.53(2)$^{1}$\\
UGC 6900                      	&	11 55 39.7	&	+31 31 07	&	7.47	&	B	&	0.7$^{+0.3}_{-0.2}$	&	9.43	&	$<$7.4	&	$<$0.26		&	8.1(3)$^{27}$\\
NGC 4068                      	&	12 04 00.8	&	+52 35 18	&	4.31	&	C	&	5		&	26.2	&	7.45(4)	&	$-$0.402(17)&	\nodata	\\
NGC 4144                      	&	12 09 58.6	&	+46 27 26	&	9.8		&	C	&	5		&	19.3	&	8.83(4)	&	0.044(4)	&	\nodata	\\
NGC 4163                      	&	12 12 09.2	&	+36 10 09	&	2.96	&	B	&	0.5$^{+0.2}_{-0}$	&	6.78	&	6.59(5)	&	$-$0.56(2)	&	7.56(14)$^{1}$\\
UGC 7267                      	&	12 15 23.7	&	+51 21 00	&	7.33	&	C	&	5		&	10.5	&	7.10(6)	&	$-$0.47(3)	&	\nodata	\\
CGCG 269$-$049                  &	12 15 46.6	&	+52 23 14	&	3.23	&	C	&	5		&	\nodata	&	$<$6.3	&	$<-$0.42	&	7.43(6)$^{46}$\\
NGC 4288                      	&	12 20 38.1	&	+46 17 30	&	7.67	&	C	&	5		&	40.5	&	8.37(4)	&	0.12(2)		&	8.5(2)$^{10}$\\
UGC 7408                      	&	12 21 15.0	&	+45 48 41	&	6.87	&	B	&	1$^{+0}_{-0.5}$		&	\nodata	&	$<$7.4	&	$<-$0.18	&	\nodata	\\
UGCA 281                      	&	12 26 15.9	&	+48 29 37	&	5.7		&	B	&	0.03$^{+0.07}_{-0}$	&	325.2	&	7.49(3)	&	$-$0.293(15)&	7.80(3)$^{47,42}$\\
								&				&				&			&	C	&	0.1$^{+0.2}_{-0}$	&			&			&				&\\
UGC 7559                      	&	12 27 05.2	&	+37 08 33	&	4.87	&	C	&	5		&	36.2	&	6.97(5)	&	$-$0.77(2)	&	\nodata	\\
UGC 7577                      	&	12 27 40.9	&	+43 29 44	&	2.74	&	B	&	0.7$^{+0}_{-0.2}$	&	8.57	&	6.60(6)	&	$-$0.69(6)	&	7.97(6)$^{1}$\\
NGC 4449                      	&	12 28 11.1	&	+44 05 37	&	4.21	&	C	&	5		&	58.5	&	9.38(3)	&	0.142(18)	&	8.31(7)$^{44,40,22,48}$\\
UGC 7599                      	&	12 28 28.6	&	+37 14 01	&	6.9		&	B	&	0.2$^{+0.3}_{-0.1}$	&	11.4	&	$<$7.1	&	$<-$0.48	&	\nodata	\\
								&				&				&			&	C	&	0.1$^{+0.1}_{-0}$	&			&			&				&\\
UGC 7605                      	&	12 28 38.7	&	+35 43 03	&	4.43	&	C	&	0.8$^{+0.2}_{-0.1}$	&	29.8	&	$<$7.56	&	$<-$0.62	&	7.66(11)$^{1}$\\
UGC 7608                      	&	12 28 44.2	&	+43 13 27	&	7.76	&	C	&	5		&	49.5	&	7.65(4)	&	$-$0.501(18)&	\nodata	\\
								&				&				&			&	B	&	0.1$^{+0}_{-0.13}$	&			&			&				&\\
NGC 4485                      	&	12 30 31.1	&	+41 42 04	&	7.07	&	C	&	5		&	66.7	&	8.77(5)	&	0.19(2)		&	\nodata	\\
UGC 7690                      	&	12 32 26.9	&	+42 42 15	&	7.73	&	C	&	5		&	21.2	&	8.06(4)	&	$-$0.157(16)&	\nodata	\\
UGC 7698                      	&	12 32 54.4	&	+31 32 28	&	6.1		&	B	&	0.7$^{+0}_{-0.2}$	&	40		&	7.44(5)	&	$-$0.57(3)	&	8.0(2)$^{27}$\\
UGC 7719                      	&	12 34 00.5	&	+39 01 09	&	9.39	&	C	&	5		&	49.5	&	7.44(3)	&	$-$0.30(2)	&	\nodata	\\
UGC 7774                      	&	12 36 22.7	&	+40 00 19	&	7.44	&	C	&	5		&	21		&	7.43(5)	&	$-$0.16(2)	&	\nodata	\\
NGC 4618                      	&	12 41 32.8	&	+41 09 03	&	7.79	&	C	&	5		&	25.6	&	9.16(4)	&	0.196(11)	&	\nodata	\\
NGC 4625                      	&	12 41 52.7	&	+41 16 26	&	8.65	&	C	&	5		&	16.1	&	8.70(4)	&	0.33(3)		&	8.4(2)$^{49}$\\
UGC 7866                      	&	12 42 15.1	&	+38 30 12	&	4.57	&	B	&	0.1$^{+0.1}_{-0}$	&	46.2	&	7.01(5)	&	$-$0.80(2)	&	\nodata	\\
NGC 4707                      	&	12 48 22.9	&	+51 09 53	&	7.44	&	B	&	0.7$^{+0}_{-0.2}$	&	22.7	&	7.57(5)	&	$-$0.46(2)	&	8.4(2)$^{27}$\\
UGC 8024                      	&	12 54 05.2	&	+27 08 59	&	4.3		&	B	&	0.07$^{+0.13}_{-0}$	&	26		&	$<$6.7	&	$<-$0.94	&	7.67(6)$^{7,50}$\\
UGC 8091                      	&	12 58 40.4	&	+14 13 03	&	2.13	&	C	&	1$^{+0}_{-0}$		&	98.1	&	5.99(5)	&	$-$0.92(2)	&	7.65(6)$^{34,35,51,37}$\\
UGCA 320                      	&	13 03 16.7	&	$-$17 25 23	&	7.24	&	C	&	1$^{+0}_{-0.3}$		&	50		&	7.71(4)	&	$-$0.801(18)&	8.1(2)$^{52}$\\
UGC 8201                      	&	13 06 24.9	&	+67 42 25	&	4.57	&	C	&	5$^{+0}_{-4}$		&	6.48	&	6.88(8)	&	$-$1.06(6)	&	7.80(6)$^{55}$\\
NGC 5023                      	&	13 12 12.6	&	+44 02 28	&	5.4		&	C	&	5		&	17.4	&	7.96(2)	&	0.042(6)	&	\nodata	\\
CGCG 217$-$018                  &	13 12 51.8	&	+40 32 35	&	8.21	&	B	&	0.5$^{+0.2}_{-0}$	&	20		&	7.72(4)	&	$-$0.150(18)&	\nodata	\\
UGC 8313                      	&	13 13 53.9	&	+42 12 31	&	8.72	&	C	&	5		&	43.4	&	8.76(4)	&	$-$0.0101(12)&	\nodata	\\
UGC 8320                      	&	13 14 28.0	&	+45 55 09	&	4.33	&	C	&	5		&	\nodata	&	7.39(7)	&	$-$0.406(19)&	8.29(7)$^{27,51}$\\
NGC 5204                      	&	13 29 36.5	&	+58 25 07	&	4.65	&	C	&	5		&	49.6	&	8.37(4)	&	$-$0.108(14)&	\nodata	\\
NGC 5264                      	&	13 41 36.7	&	$-$29 54 47	&	4.53	&	B	&	0.5$^{+0.4}_{-0}$	&	7.63	&	7.75(9)	&	$-$0.09(2)	&	8.7(2)$^{52}$\\
UGC 8760                      	&	13 50 50.6	&	+38 01 09	&	3.24	&	C	&	5		&	11.7	&	$<$6.6	&	$<-$0.44	&	\nodata	\\
NGC 5477                      	&	14 05 33.3	&	+54 27 40	&	7.7		&	C	&	0.8$^{+0.2}_{-0.1}$	&	54.2	&	7.72(4)	&	$-$0.47(2)	&	8.14(7)$^{53}$\\
UGC 9128                      	&	14 15 56.5	&	+23 03 19	&	2.24	&	B	&	0.7$^{+0}_{-0.2}$	&	3.92	&	$<$6.1	&	$<-$0.43	&	7.75(5)$^{7,8,29}$\\
NGC 5585                      	&	14 19 48.2	&	+56 43 45	&	5.7		&	C	&	5		&	24.6	&	8.50(4)	&	$-$0.11(2)	&	\nodata	\\
								&				&				&			&	B	&	0.3$^{+0.2}_{-0}$	&			&			&				&\\
UGC 9240                      	&	14 24 43.4	&	+44 31 33	&	2.8		&	C	&	5		&	8.26	&	6.85(4)	&	$-$0.519(18)&	7.95(3)$^{9,51}$\\
IC 5052                       	&	20 52 05.6	&	$-$69 12 06	&	5.86	&	B	&	0.5$^{+0.2}_{-0}$	&	39.7	&	8.67(4)	&	0.371(12)	&	\nodata	\\
NGC 7064                      	&	21 29 03.0	&	$-$52 46 03	&	9.86	&	C	&	5		&	\nodata	&	8.34(4)	&	$-$0.368(9)	&	\nodata	\\	
UGC 12613                     	&	23 28 36.3	&	+14 44 35	&	0.76	&	B	&	0.7$^{+0}_{-0.1}$		&	0.98	&	5.93(5)	&	0.057(8) 	&	7.93(14)$^{54}$\\
UGCA 442                      	&	23 43 45.6	&	$-$31 57 24	&	4.27	&	C	&	5		&	23.6	&	6.85(5)	&	$-$0.77(5)	&	7.72(3)$^{13,15,52}$
\enddata
\tablecomments{
Columns list the 
(1) Galaxy name,
(2) Right Ascension,
(3) Declination in J2000 coordinates,
(4) Luminosity distance in Mpc,
(5) Instantaneous burst (B) or constant (C) SFH of the best-fit model to observations,
(6) Age of the best-fit model in Gyr, acquired with SED fitting with the 1$\sigma$ range,
(7) H$\alpha$ equivalent width (\AA), from \citet{kennicutt08}, 
(8) log of the total integrated IR luminosity per solar luminosity $L_{\sun}$, 
(9) IRX values of $\log L_{\rm TIR}/L_{\rm FUV}$, and 
(10) 12+log(O/H) oxygen abundance values for each galaxy (when available) and respective references.   Numbers in parentheses indicate uncertainties in the final digit(s) of listed quantities, when available.  In some cases, a galaxy SED can be best-fit with both a constant SFR and an instantaneous burst of star formation.  We have listed both the constant and bursting model for these galaxies.  An age notation of 0.5$^{+0.4}_{-0}$~Gyr implies a best-fit age of 500~Myr with a 1$\sigma$ spread in the age ranging from 500 to 900~Myr.  For the constant SFR case, the 5~Gyr duration is generally indicated without error bars, because this is the maximum age value we fit.  Galaxies may have longer duration.  \\
Oxygen abundance references:  
(1) -- \citet{moustakas10}; 
(2) -- \citet{tullmann03}; 
(3) -- \citet{zaritsky94}; 
(4) -- \citet{saviane08}; 
(5) -- \citet{christensen97}; 
(6) -- \citet{vilacostas93}; 
(7) -- \citet{vanzee97a}; 
(8) -- \citet{vanzee97b}; 
(9) -- \citet{vanzeeh06}; 
(10) -- \citet{kewley05}; 
(11) -- \citet{magrini07}; 
(12) -- \citet{rosolowsky08}; 
(13) -- \citet{skillman03}; 
(14) -- \citet{hidalgo01}; 
(15) -- \citet{miller96}; 
(16) -- \citet{walsh97}; 
(17) -- \citet{raimann00}; 
(18) -- \citet{aguero97}; 
(19) -- \citet{storchi94}; 
(20) -- \citet{masegosa94}; 
(21) -- \citet{lee04}; 
(22) -- \citet{hunter82}; 
(23) -- \citet{gildepaz00a}; 
(24) -- \citet{gildepaz00b}; 
(25) -- \citet{kniazev04}; 
(26) -- \citet{izotov06}; 
(27) -- \citet{hunter99}; 
(28) -- \citet{pustilnik03}; 
(29) -- \citet{skillman89};  
(30) -- \citet{kinman81}; 
(31) -- \citet{hopp91};  
(32) -- \citet{pustilnik05}; 
(33) -- \citet{hunter85}; 
(34) -- \citet{vanzee06}; 
(35) -- \citet{moles90};  
(36) -- \citet{lee05};  
(37) -- \citet{stasinska86}; 
(38) -- \citet{millerh96};  
(39) -- \citet{moustakas06}; 
(40) -- \citet{mccall85};  
(41) -- \citet{guseva00}; 
(42) -- \citet{thuan05}; 
(43) -- \citet{buckalew05}; 
(44) -- \citet{martin97}; 
(45) -- \citet{gallagher89}; 
(46) -- \citet{kniazev03};  
(47) -- \citet{perezmontero03}; 
(48) -- \citet{kibulnicky99};  
(49) -- \citet{gildepaz07};  
(50) -- \citet{kennicutt01}; 
(51) -- \citet{hidalgo02};  
(52) -- \citet{lee03};  
(53) -- \citet{izotov07}; 
(54) -- \citet{skillman97};
(55) -- \citet{berg12};
(56) -- \citet{croxall09}.
}
\end{deluxetable*}

\section{Definitions}\label{sec:3}
\subsection{UV Colors}\label{sec:UVcolor}
The horizontal axis of the IRX-$\beta$ diagram was originally the slope $\beta$ of the UV spectrum: f$_{\lambda}\propto\lambda^{\beta}$, as defined by \citet{calzetti94}.  Later, \citet{kong04} proposed a re-definition based on the FUV and NUV GALEX bands to facilitate analysis using the large dataset provided by this mission: 
\begin{equation}\label{eq:1}
\beta_{\rm GLX} = \frac{\log f_{\rm FUV} - \log f_{\rm NUV}}{\log \lambda_{\rm FUV} - \log \lambda_{\rm NUV}},
\end{equation} 
where $f_{\rm FUV}$ and $f_{\rm NUV}$ (erg~s$^{-1}$~cm$^{-2}$~\AA$^{-1}$) are the flux density per unit wavelength of the FUV ($\lambda_{eff}=1520$~\AA) and NUV ($\lambda_{eff}=2310$~\AA) bands, respectively.  Here we adopt the $\beta_{\rm GLX}$ defined by \citet{kong04}, for homogeneity with previous works.

\subsection{The IRX-$\beta$ Relation}
In the original definition of \citet{meurer99}, the IRX $=\log(L_{\rm IR}/L_{\rm FUV})$ included, in the
calculation of $L_{\rm IR}$, only the dust luminosity in the range 42--122 \micron.  The denominator is the observed FUV luminosity.  In the re-definition of \citet{kong04}, which we use here, the dust luminosity, $L_{\rm TIR}$, is the integrated emission in the 8--1000 \micron\ range, while the denominator is still the observed FUV luminosity:
\begin{equation}\label{eq:2}
{\rm IRX} \equiv \log \left(\frac{L_{\rm TIR}}{L_{\rm FUV}}\right),
\end{equation}
where $L_{\rm TIR}$ is calculated from the MIPS 24, 70, and 160~\micron\ band $Spitzer$ observations, using the recipe of \citet{dale02}.  $L_{\rm FUV}$ is from the GALEX shortest wavelength photometry.  We assume that the stellar light lost in the UV/optical regime to dust is fully recovered in the IR as dust emission, implying that the galaxy average dust scattering is negligible. 

For sources that have any combination of upper limits for the flux density of 24, 70, or 160~$\micron$, we treat the estimate of the IR luminosity as an upper limit.  The calculated luminosities and their uncertainties are listed in Table \ref{tab:2}.

\subsection{The IRX-$\beta$ Diagram}
\begin{figure}
\epsscale{1.2}
\plotone{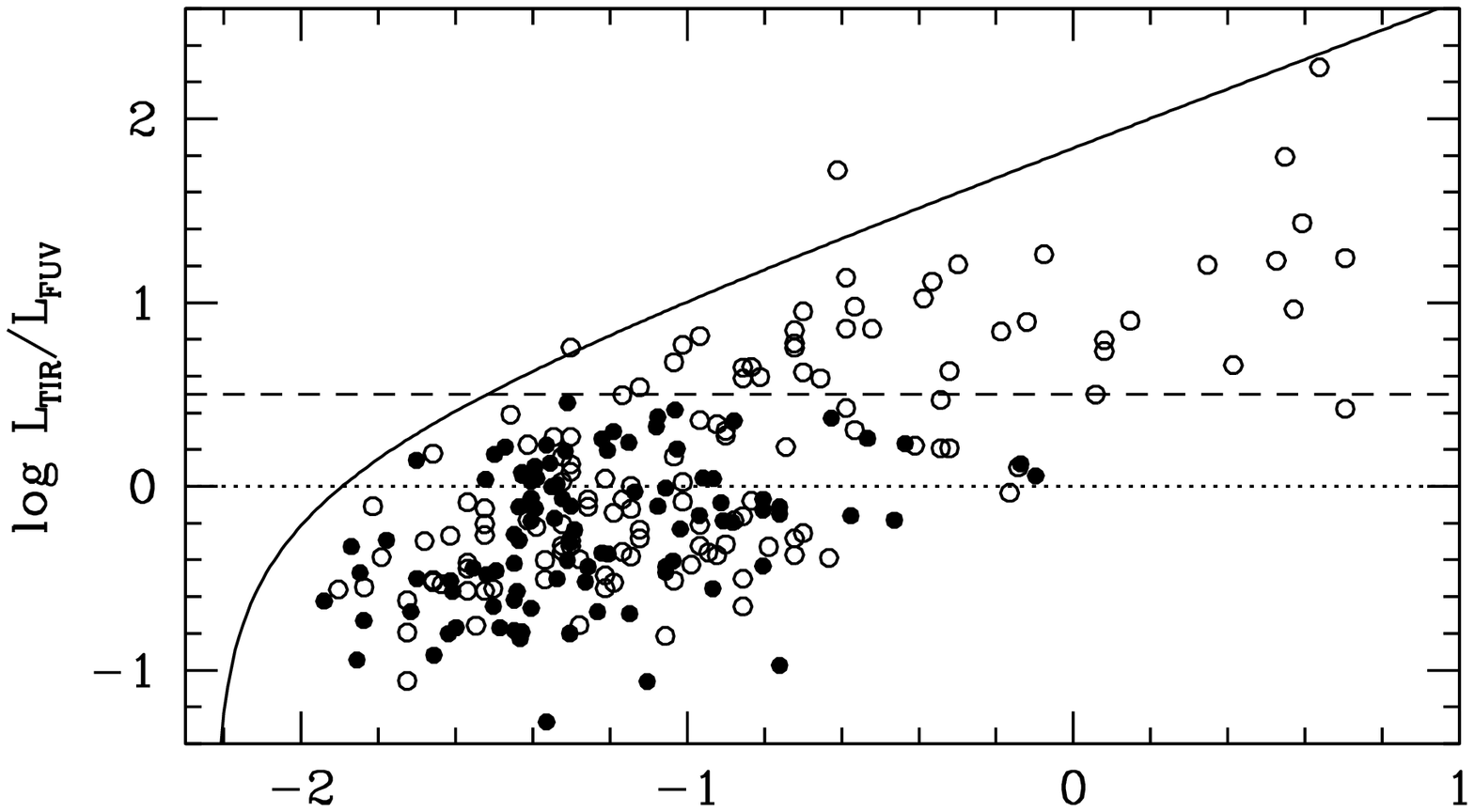}
\plotone{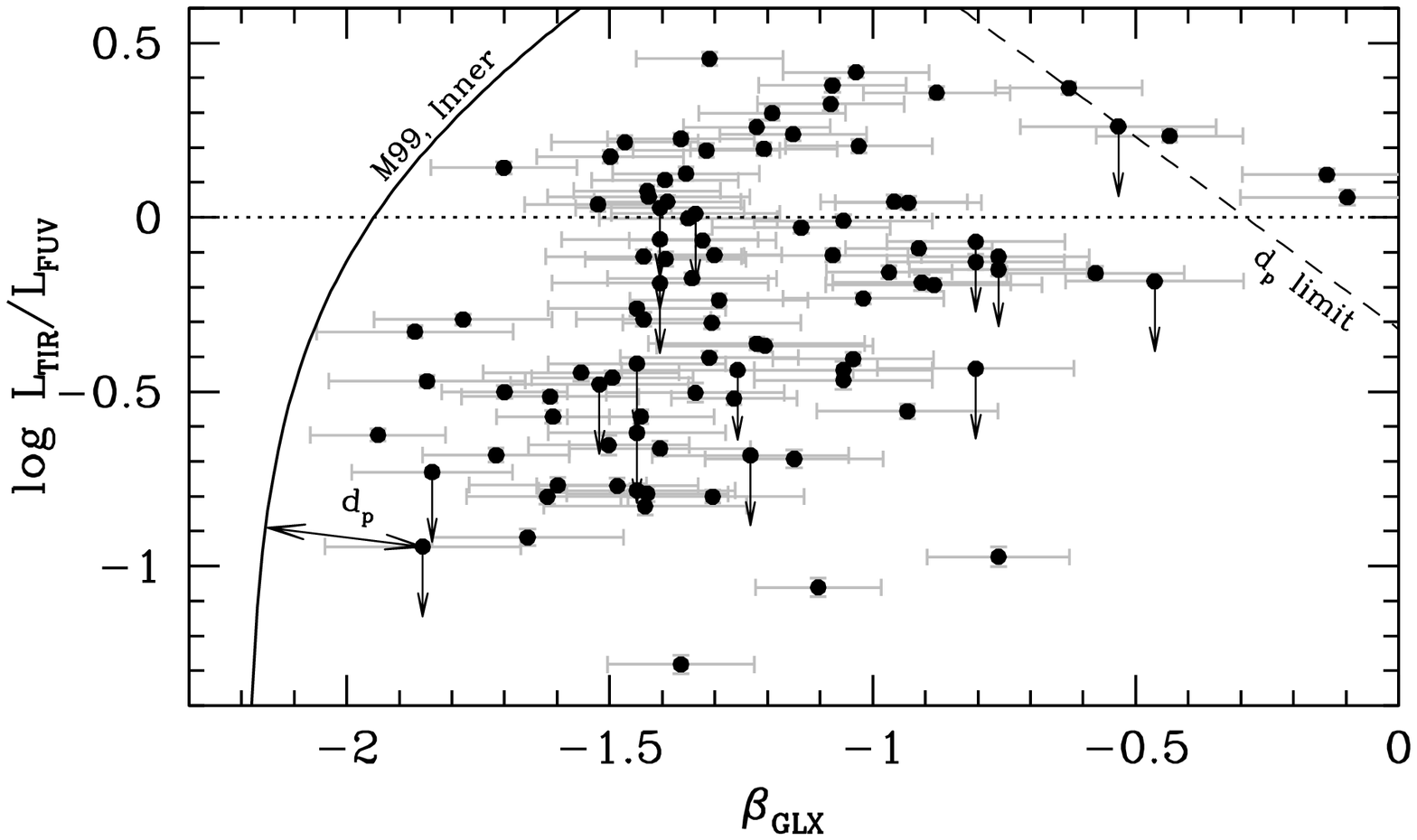}
\caption{$Top:$  The IRX-$\beta$ diagram, showing the ratio log $L_{\rm TIR}/L_{\rm FUV}$ as a function of the UV color, $\beta_{\rm GLX}$, for the entire sample of LVL galaxies.  The solid black circles below the dashed line at IRX $=0.5$ indicate the sub-set of galaxies with the required photometry to make it in our sample.  The dotted line at IRX $=0$ (top and bottom panels) indicates the cut-off for the galaxies that we consider dust-free.  The solid black line shows the starburst IRX-$\beta$ relation, determined by a least-squares fit to starburst galaxies as revised by \citet{overzier11}.  $Bottom$:  The IRX-$\beta$ diagram for our sample of 98 normal star-forming galaxies.  All downward pointing arrows represent galaxies with upper limit values of the total bolometric IR luminosity, $L_{\rm TIR}$.  The perpendicular distance $d_p$ represents the shortest distance between a galaxy and the starburst attenuation curve.  The dotted line at IRX $=0$ indicates the cut-off for the galaxies that we consider dust-free.  The angled dashed line in the upper right-hand corner marks the completeness region for sources with $d_p$ analysis; the three sources that lie rightward are excluded from all perpendicular distance results.  The gray bars represent the 1$\sigma$ errors. 
\label{fig:1}}
\end{figure}

As we want to compare our normal star-forming galaxies to the starbursts, we adopt as our reference IRX-$\beta$ relation the re-derivation of the \citet{meurer99} relation by \citet{overzier11}.  These authors consider the case in which both the UV and IR apertures only include the central starburst regions of galaxies, thus deriving a `pure starburst' relation which is close to the original one of \citet{meurer99}.  The starburst relation from \citet{overzier11}, which they call `M99, Inner', reported to our conventions of equations \ref{eq:1} and \ref{eq:2}, is:
\begin{equation}\label{eq:3}
\log {\rm IRX} = \log (10^{0.4(4.54+2.07\beta_{\rm GLX})}-1) + 0.225, 
\end{equation}
and is plotted in Figure \ref{fig:1} together with the 98 galaxies that form our sample.  As already known, our galaxies show a large scatter and IR/UV values that are systematically below Eq. \ref{eq:3}.  We further segregate our sample into the 68 galaxies located below $\log L_{\rm TIR}/L_{\rm FUV} = 0$, which we call ``dust-free'' galaxies, and the 30 ``slightly-dusty'' galaxies at $0 < \log L_{\rm TIR}/L_{\rm FUV} \leq 0.5$ (Figure \ref{fig:1}).  We make this distinction to investigate degeneracies that may be introduced by increasing the dust contribution to the SED of a galaxy.

\subsection{Perpendicular Distance $d_p$}
In order to quantify the deviation of a galaxy from the starburst relation in the IRX-$\beta$ diagram, \citet{kong04} introduced the concept of `shortest distance' or `perpendicular distance' $d_p$ of a galaxy's position in this diagram from the IRX-$\beta$ starburst relation: 
\begin{equation}\label{eq:4}
d_p = \sqrt{(x_i-x_{\rm SB})^2+(y_i-y_{\rm SB})^2},
\end{equation}
where $(x_i,y_i)$ are the $\beta_{\rm GLX}$ IRX coordinates of an individual galaxy, $(x_{\rm SB}, y_{\rm SB})$ are the $\beta_{\rm GLX}$ and IRX values of the starburst curve that is closest to the galaxy coordinates (Figure \ref{fig:1}).  $d_p$ is a dimensionless quantity and we will be following the convention of \citet{kong04} by assigning a positive $d_p$ distance to galaxies that have values of $L_{\rm TIR}/L_{\rm FUV}$ lying above the starburst IRX-$\beta$ relation and a negative $d_p$ to galaxies that lie below the starburst relation.  In our specific case, all $d_p$ values are negative.  In order to guarantee complete sampling and sound statistics of any correlation with $d_p$, we will exclude from our analysis three galaxies that are above the dashed line drawn in the bottom panel of Figure \ref{fig:1}, the $d_p$ limit.  This ensures statistical completeness along the $d_p$ axis for all galaxies within our selection criteria (IRX $\le0.5$).

\section{Modeling}\label{sec:4}
We derive ages for our galaxies by comparing their FUV-to-Ks band SEDs with spectral synthesis models of stellar populations, modulated by (some) dust attenuation, via $\chi^2$ minimization.  As age and extinction are degenerate, we use the FIR data to constrain the total amount of stellar light that can be absorbed by dust and re-processed in the infrared, in order to decrease the number of valid age/extinction solutions our best-fitting routine yields.  Details on the procedure are given below.

\subsection{Generating the Synthetic Spectral Energy Distributions}\label{sec:makingSEDs}
We generate synthetic models from Starburst99 \citep{leitherer99}, with metallicity values of $Z = 0.0004$, 0.004, 0.008, and 0.020, and with a Kroupa stellar initial mass function (IMF) from 0.1~M$_{\sun}$ to 100 M$_{\sun}$, with exponents of 1.3 in the mass range of $0.1-0.5$~M$_{\sun}$ and 2.3 over the mass range $0.5-100$~M$_{\sun}$.  For galaxies with known metal abundance values, we adopt the synthetic models with the closest metallicity value to the measured one.   We adopt the Padova stellar evolutionary tracks, with and without the thermally pulsating asymptotic giant branch (AGB).  A posteriori, we verify that the stellar evolutionary track has little influence on the results:  no galaxy favored the original Padova over the AGB-Padova track.  The SED models cover the age range 10 Myr to 5 Gyr, for both constant star formation (C) and for an instantaneous burst (B).  The actual SFH of a galaxy will be far more complex than the two simple ones we implement (C or B), and may exhibit a succession of bursts or increasing/decreasing SFH, etc.  Our approach, however, covers the extremes in terms of behavior of the UV colors \citep{calzetti05}, and we expect them to bracket the range of observed colors.  We do not aim at determining accurate SFHs, but simply to `bin' our galaxies, according to this coarser classification. 

Luminosities and colors are derived by convolving the generated synthetic SEDs with the spectral response function of all our filters.  The convolved fluxes are then compared with the observed fluxes.  Dust attenuation (see next section) is applied to the SEDs prior to the filter band-pass convolution.  In addition to filter-convolved SEDs, we use Starburst99 to derive the expected values of the: FUV/NIR ratio, U$-$B colors, and \ewha\ for each age/metallicity/SFH combination, which we will compare against {\em extinction--corrected} observational data.

\subsection{Dust Attenuation Models}\label{sec:attmodels}
Although our galaxies are selected to have low-dust content, we still implement dust corrections in our fitting routines, in order to recover the intrinsic age/duration.  We use, as default, the starburst attenuation curve \citep{calzetti00}, and probe the range of color excess $E(B-V)=$ [0.0,0.3].  As we will discuss later, our galaxies have best-fit color excesses $E(B-V)\lesssim0.08$, in agreement with the requirement that they are virtually dust-free.  As a test of the robustness of our results, we check the effect of using the Small Magellanic Cloud \citep[SMC;][]{bouchet85} extinction curve with both foreground and homogeneously-mixed geometries \citep{calzetti01}.  We find variations in the age determinations $\lesssim$100~Myr for $E(B-V)=0.07$, consistent with our typical uncertainty (Table \ref{tab:2}).

\subsection{Fitting the UV-Optical SED}
\begin{figure}
\epsscale{1.2}
\plotone{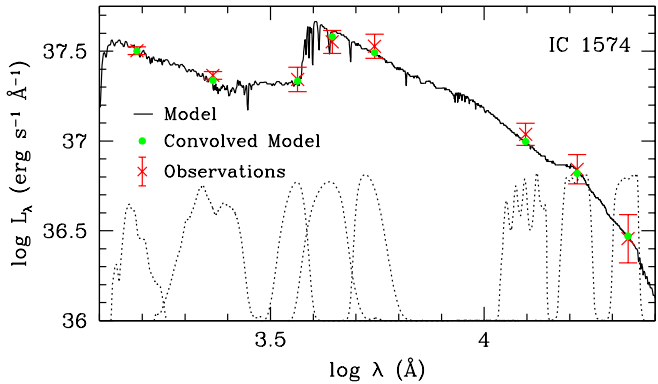}
\caption{A best-fit Starburst99 SED from the FUV to the Ks bands for IC 1574.  The black spectrum represents the synthetic Starburst99 spectrum, the red points represent the galaxy observations with accompanying 1$\sigma$ errors, the eight dotted lines represent the filter transmission curve for the FUV, NUV, U, B, V, J, H, and Ks bands, and the solid green circles represent the convolved flux values of the Starburst99 curve with each transmission curve.  The value of $\chi^2_{\rm min} = 0.5$.  
\label{fig:2}}
\end{figure}

For each galaxy in our sample, the best-fit models are determined with a $\chi^2$ minimization:
\begin{equation}\label{eq:5}
\chi^2_i= \sum_{bp} \left( \frac{F_{{\rm obs},bp} - c_i F_{{\rm mod_i},bp}}{\sigma(F_{{\rm obs},bp})}  \right)^2,
\end{equation}
where $i$ represents the $i^{\rm th}$ model SED and we sum over the band-passes ($bp$) FUV, NUV, U, B, V, J, H, and Ks, giving us seven degrees of freedom in determining the scale factor $c_i$:
\begin{equation}\label{eq:6}
c_i = \sum_{bp} \left( \frac{F_{{\rm obs},bp}F_{{\rm mod_i},bp}}{\sigma^2(F_{{\rm obs},bp})}\right) \Big/ \sum_{bp} \left( \frac{F_{{\rm obs},bp}}{\sigma(F_{{\rm obs},bp})}  \right)^2.
\end{equation}
We exclude from the fits all data long-ward of Ks, to avoid even marginal dust emission contamination of the stellar fluxes.  $F_{{\rm obs},bp}$ is the observed flux at each band-pass, $\sigma(F_{{\rm obs},bp})$ are accompanying $1\sigma$ errors, and $F_{{\rm mod_i},bp}$ is the flux of each individual model, first dust attenuated and then convolved with the filter band-pass (Section \ref{sec:makingSEDs} and \ref{sec:attmodels}).  An example is given in Figure \ref{fig:2}.

The $\chi^2_i$ value for each model is then assigned a weight $w_i=exp(-\chi^2_i/2)$, giving a probability distribution function (PDF) for the galaxy parameters of interest.  We accept as valid all models that lie within a factor of three from $\chi^2_{\rm min}$, and average the resulting parameters of interest: age (duration for the case of constant star formation), metallicity, color excess $E(B-V)$, and SFH (C or B).  This process ensures that we have not selected a single model that fits the observations by chance.  With this requirement, there are often times an arbitrarily large number of acceptable models that all provide reasonable fits to the data.  We, thus, use conservation of energy to further constrain the number of acceptable models.  The total amount of light absorbed by dust in the range $912-22,000$~\AA\ has to equal the aggregate emission in the infrared ($L_{\rm TIR}$) within a factor of two.

In order to test the robustness of our fitting procedure in regard to the assigned SFH (C or B) to each galaxy, we repeat the procedure above, but accept all models that are within a factor of five from $\chi^2_{\rm min}$.  Even with this more relaxed criterion, no galaxy undergoes a switch in the SFH; their designation (C or B) remains unchanged, although the number of acceptable solutions increases, as expected.  We thus conclude that the SFH assignment for each galaxy is a robust result, albeit within the limits of the simplified SFHs we explore.  For galaxies best-fit by constant star formation, the more relaxed criterion did not translate in a larger range of acceptable parameter values (duration, $E(B-V)$, metallicity).  For galaxies best-fit by instantaneous bursts the range increased; however, in the case of the best-fit age, the increase is never larger than 100~Myr, which is within our typical uncertainty.  Because of the limited effect of changing the threshold for acceptable fits, we choose our original selection criterion of keeping all best-fit model solutions within a factor of three from $\chi^2_{\rm min}$, and use the spread in the resulting parameters to assign the 1$\sigma$ uncertainty to our results.  Table 2 lists the results of our fitting procedure.

Our $\chi^2$ routine is performed using YAFIT in the Java programming language.  Six galaxies are well fit with both bursting and constant star forming models; these galaxies have both models listed in Table 2, but we only show the bursting model in all of our plots.

\section{Results and Analysis}\label{sec:5}
\subsection{Mean Best-Fit Age of the Stellar Populations}
\begin{figure}
\epsscale{1.2}
\plotone{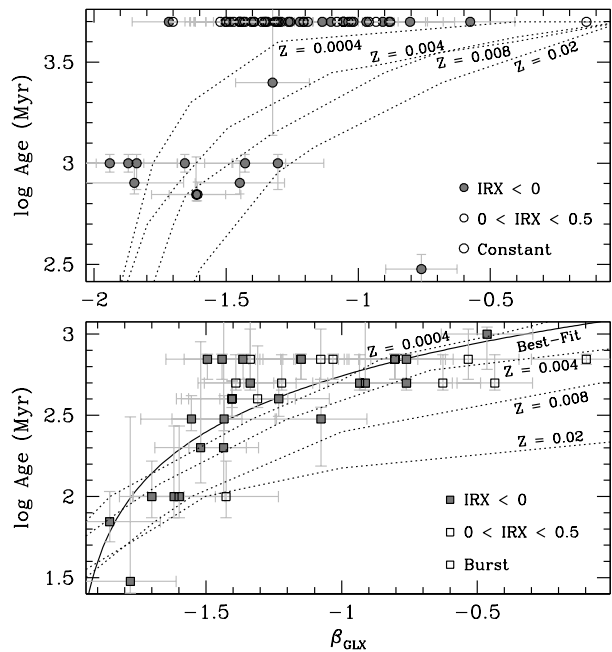}
\caption{$Top$:  The mean age of each galaxy as a function of $\beta_{\rm GLX}$ for galaxies that are best-fit with constant SFRs (circles).  Open symbols represent galaxies with an IRX value of $0< \log L_{\rm TIR}/L_{\rm FUV} \leq 0.5$ and filled symbols represent galaxies with an IRX value of $\log L_{\rm TIR}/L_{\rm FUV} \leq 0$.  The vertical error bars represent the range of acceptable model ages for each galaxy.  The lines are predicted trends for the age as a function of $\beta_{\rm GLX}$ by Starburst99 models for constant star-forming (dotted) systems at metallicities of $Z=0.0004, 0.004, 0.008$ and 0.02.  
$Bottom$:  The age of each galaxy as a function of $\beta_{\rm GLX}$ for galaxies that are best-fit with bursting SFRs (squares).  The solid black line represent the least-squares fit to the entire sample of bursting galaxies.  The dotted lines are the predicted trends for the age as a function of $\beta_{\rm GLX}$ from Starburst99 models for burst galaxies.
\label{fig:agebeta}}
\end{figure}

\begin{figure}
\epsscale{1.2}
\plotone{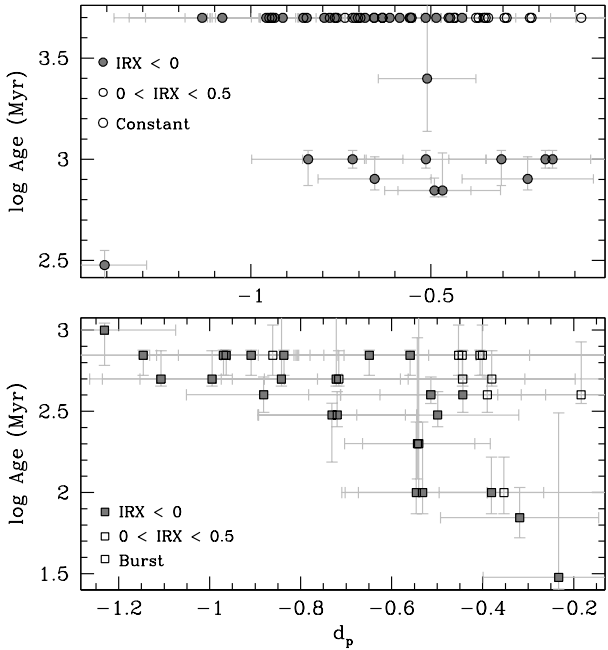}
\caption{$Top$:  The mean age of each galaxy as a function of perpendicular distance $d_p$ for galaxies that are best-fit with a constant SFR (circles).  The symbol are the same as in Figure \ref{fig:agebeta}.  The vertical error bars represent the range of acceptable model ages for each galaxy.  
$Bottom$:  The age of each galaxy as a function of perpendicular distance $d_p$ for galaxies that are best-fit with a bursting SFR (squares).  
\label{fig:agedp}}
\end{figure}

The mean age of each galaxy -- as determined from the $\chi^2$ fitting to our models -- is shown as a function of the UV color $\beta_{\rm GLX}$ and of the perpendicular distance $d_p$ in Figures \ref{fig:agebeta} and \ref{fig:agedp}, respectively.  We also list both the Spearman rank $\rho$ and Kendall $\tau$ correlation coefficients\footnote{Spearman $\rho$ is a non-parametric product-moment correlation coefficient that measures the statistical dependency between two variables computed with the rank of the data.  Kendall $\tau$ is a non-parametric hypothesis test that measures the association between two measured quantities and represents a probability (i.e., the difference between the probability that the observed data are in the same order versus the probability that the observed data are not).} as determined for nearly every variable in our study as a function of both $\beta_{\rm GLX}$ and $d_p$ in Tables 3, 4, and 5 for instantaneous burst galaxies, constant star-forming galaxies, and the entire sample, respectively.  In the Tables, we separate the samples into subsets on the IRX-$\beta$ diagram, performing statistics on the galaxies below and above $\log L_{\rm TIR}/L_{\rm FUV} = 0$.  We use both correlation coefficients as they have different interpretations.  Usually the values returned by both statistics are very similar; when discrepancies occur, we adopt the lower (i.e., less significant) value.

In both Figures \ref{fig:agebeta} and \ref{fig:agedp} we observe that galaxies better described by an instantaneous burst (`B-galaxies' from now on) show a stronger trend with both $\beta_{\rm GLX}$ and $d_p$ than galaxies better described by constant star formation (`C-galaxies' from now on).  The visual impression is supported by the correlation coefficient analysis:  we find that the correlation between age and $\beta_{\rm GLX}$ has a significance between 4$\sigma$ and 5$\sigma$ for B-galaxies, and is stronger than that between age and $d_p$.  C-galaxies are not only mostly uncorrelated with both parameters, but show a build up at 5 Gyr, the oldest age value we consider.  This build up is likely artificial, and simply means that most of our C-galaxies have star formation lasting over timescales longer than 5 Gyr.
 
The expectation from Starburst99 for the trend between age and $\beta_{\rm GLX}$ is shown in Figure \ref{fig:agebeta}, for all four of model Starburst99 metallicity values: $Z=0.0004$ ($\sim$1/30 solar), 0.004, 0.008, and 0.02 ($\sim$1.4 solar).  This is roughly the same range observed for our sample galaxies (see next section), which has median value 1/4th solar.  No corresponding model expectation can be reported for the the perpendicular distance d$_p$, as this quantity is not directly measurable from synthetic stellar population SEDs.

The trend marked by the model predictions for B-galaxies is in remarkable agreement with the observed trend for the Age-$\beta_{\rm GLX}$ (bottom panel of Figure~\ref{fig:agebeta}), and is better than what one could tentatively expect given the coarseness of our SFH bins (B or C).  The low-metallicity model trend overlaps almost entirely with the best-fitting curve to the data, which we determine with Levenberg-Marquandt algorithm for non-linear least squares optimization:
\begin{equation}\label{eq:7}
\log \rm{Age}(\beta_{\rm GLX}) = \log[10^{4.4\times10^{-4} + 2.3\times10^{-4}\beta_{\rm GLX} }-1]+6.1,
\end{equation}
and is shown in the bottom panel of Figure \ref{fig:agebeta}.  The line in Eq. \ref{eq:7} is bracketed by the models at $Z=0.0004$ and $Z=0.004$ (i.e., $\sim$1/30 to 1/3 solar), in agreement with the range of metallicity values of most of our bursting galaxies (see Figure \ref{fig:met} below).

As already remarked above, the correlation between age and $d_p$ is usually weak or non-significant, with the only exception of B-galaxies with IRX $\leq 0$, for which we find a 4$\sigma$ significance (Table 3).

\subsection{The Color Excess $E(B-V)$ and the Metallicity}
\begin{figure}
\epsscale{1.2}
\plotone{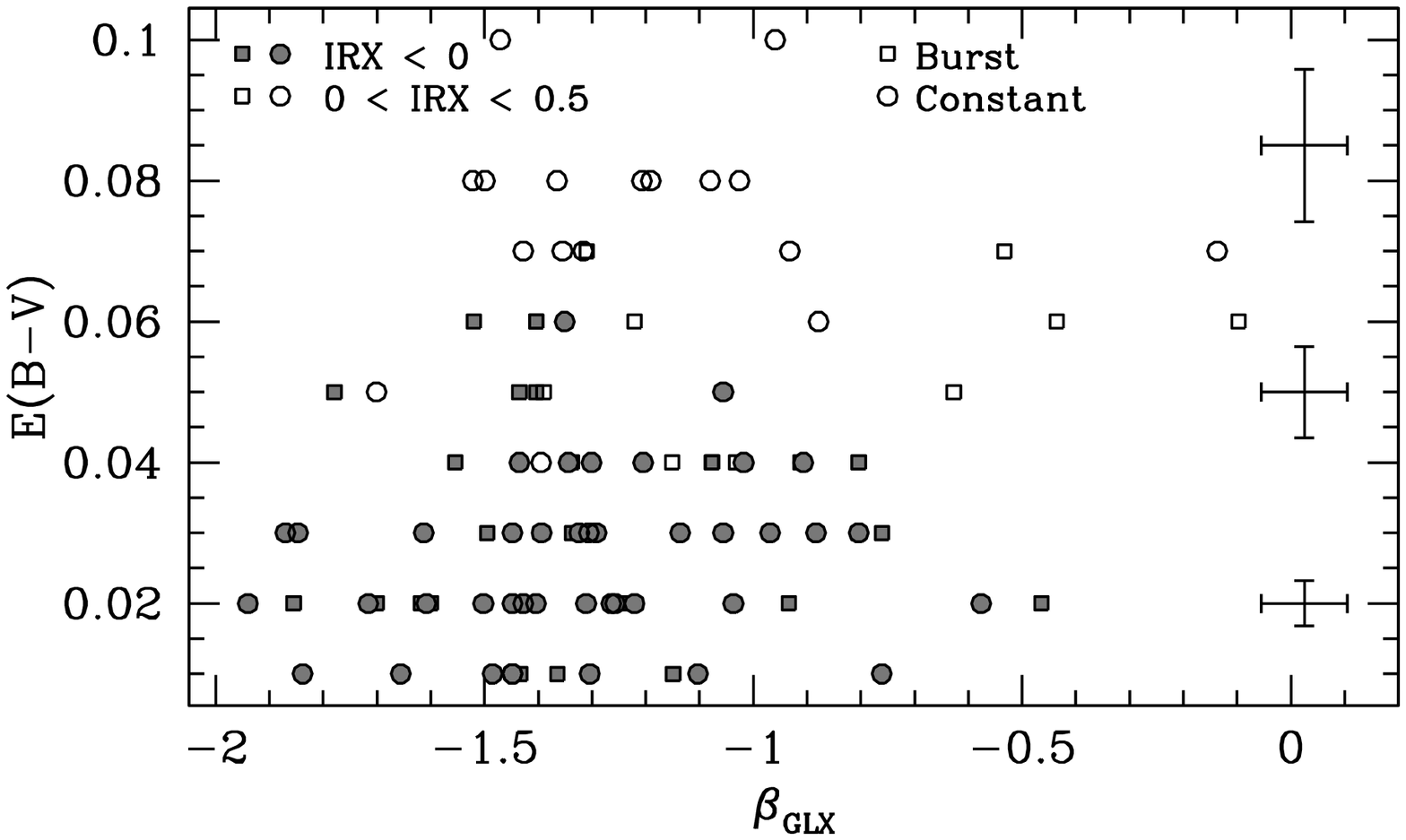}
\plotone{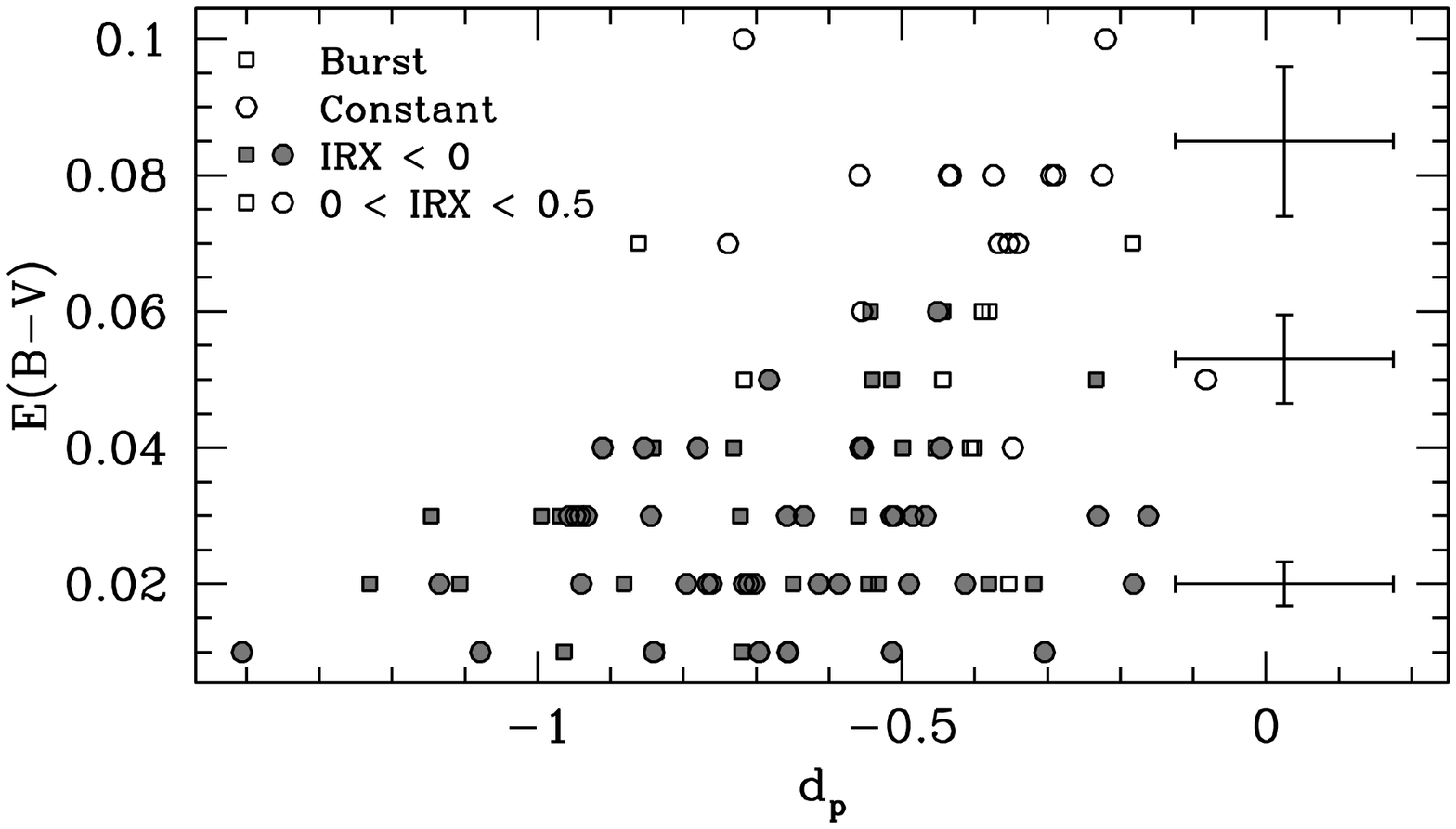}
\caption{$Top$:  The color excess $E(B-V)$ of the best-fit model for each galaxy as a function of the UV color $\beta_{\rm GLX}$.  The symbol are the same as in Figure \ref{fig:agebeta}.  The three error bars indicate the mean 1$\sigma$ uncertainty error, where the error in both $\beta$ and $d_p$ is constant but the error in $E(B-V)$ increases with increasing color excess.
$Bottom$:  The color excess $E(B-V)$ of the best-fit model for each galaxy as a function of the perpendicular distance $d_p$.  
\label{fig:E}}
\end{figure}

The extinction to be applied to the stellar population models necessary to reproduce the observations is one of the outputs of our fits, and is expressed as the color excess $E(B-V)$.  Figure \ref{fig:E} shows $E(B-V)$ as a function of $\beta_{\rm GLX}$ and $d_p$.  As expected, galaxies in the range $0<$ IRX $\leq 0.5$ have a higher average $E(B-V)$ value than galaxies with IRX $\leq 0$.  We find no correlation between the color excess of the best-fit model of each galaxy to the UV color $\beta_{\rm GLX}$, as may be expected for our low-extinction systems:  our galaxies are attenuated at most by a factor of 1.8 in the FUV.  A weak trend is observed between $E(B-V)$ with $d_p$ (Figure \ref{fig:E}, bottom panel) with an increase in the color excess as a galaxy lies closer to the starburst IRX-$\beta$ relation.  This is in the opposite direction of what one would expect if the deviations from the IRX-$\beta$ relation were driven by increased dust attenuation.  However, none of these trends is statistically significant (Tables 3, 4, and 5).

Oxygen (12+log(O/H)) abundance measurements are available for 61\% of the galaxies in our sample (60/98) from the literature, listed in Table \ref{tab:2}.  Since metallicity is a parameter we allow to vary in the Starburst99 spectra, we require that the models match the galaxy's measured metal abundance.  If there is no metallicity known a priori, the metallicity of the synthetic SEDs are allowed to accept any metallicity between $Z = 0.0004$ to 0.02.  We elect not to estimate metallicity values for galaxies without a directly measured nebular oxygen abundance.  Typical estimates (e.g., mass-metallicity relation) have a 2$\sigma$ scatter that is only a factor of two smaller than our full metallicity range \citep{tremonti04}.  The metallicity values inferred would thus have too much uncertainty to yield useful information.  Over 93\% of the galaxies in our sample with known metallicity lie below the solar metallicity value, in agreement with the expectation that the dust-poor LVL galaxies are also generally metal-poor.

\begin{figure}
\epsscale{1.2}
\plotone{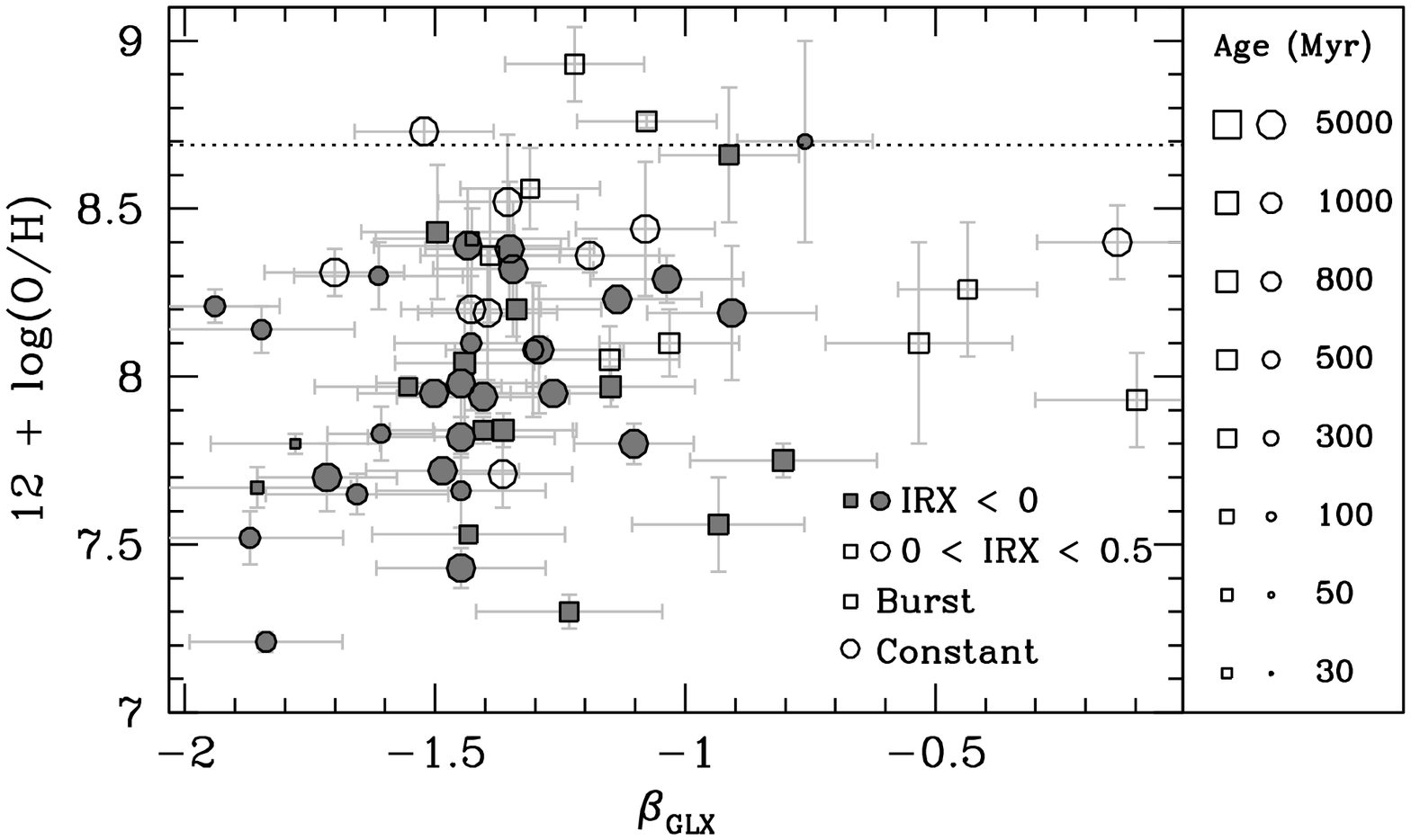}
\plotone{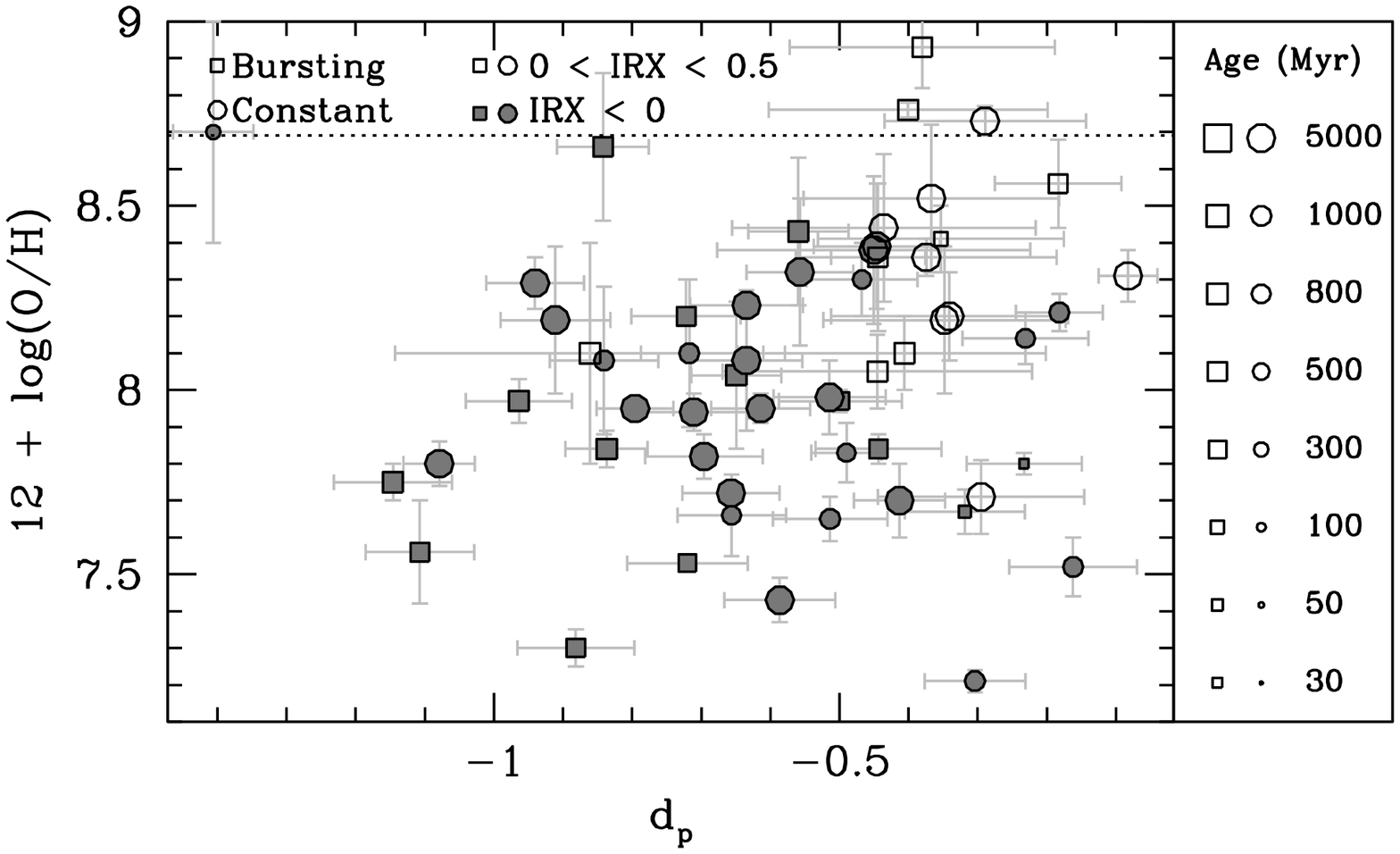}
\caption{$Top$:  Oxygen abundance measurements as a function of the UV color $\beta_{\rm GLX}$.  The squares represent galaxies that are best-fit with bursts and circles represent galaxies that are best-fit with constant SFRs.  The size of the circles/squares represents the age for the best-fit models, shown in the right-hand panel.  Open symbols represent galaxies with an IRX value of $0< \log L_{\rm TIR}/L_{\rm FUV} \leq 0.5$ and filled symbols represent galaxies with an IRX value of $\log L_{\rm TIR}/L_{\rm FUV} \leq 0$.  The gray bars represent the 1$\sigma$ errors.  The dotted horizontal line represents the solar metallicity value. 
$Bottom$:  12+log(O/H) measurements as a function of the perpendicular distance $d_p$.  
\label{fig:met}}
\end{figure}

As for the color excess, we find no significant correlation between the metal content of our galaxies and either $\beta_{\rm GLX}$ or $d_p$ (Figure \ref{fig:met}). 

\subsection{Other Estimators of Mean Stellar Population Age}
\begin{figure}
\epsscale{1.2}
\plotone{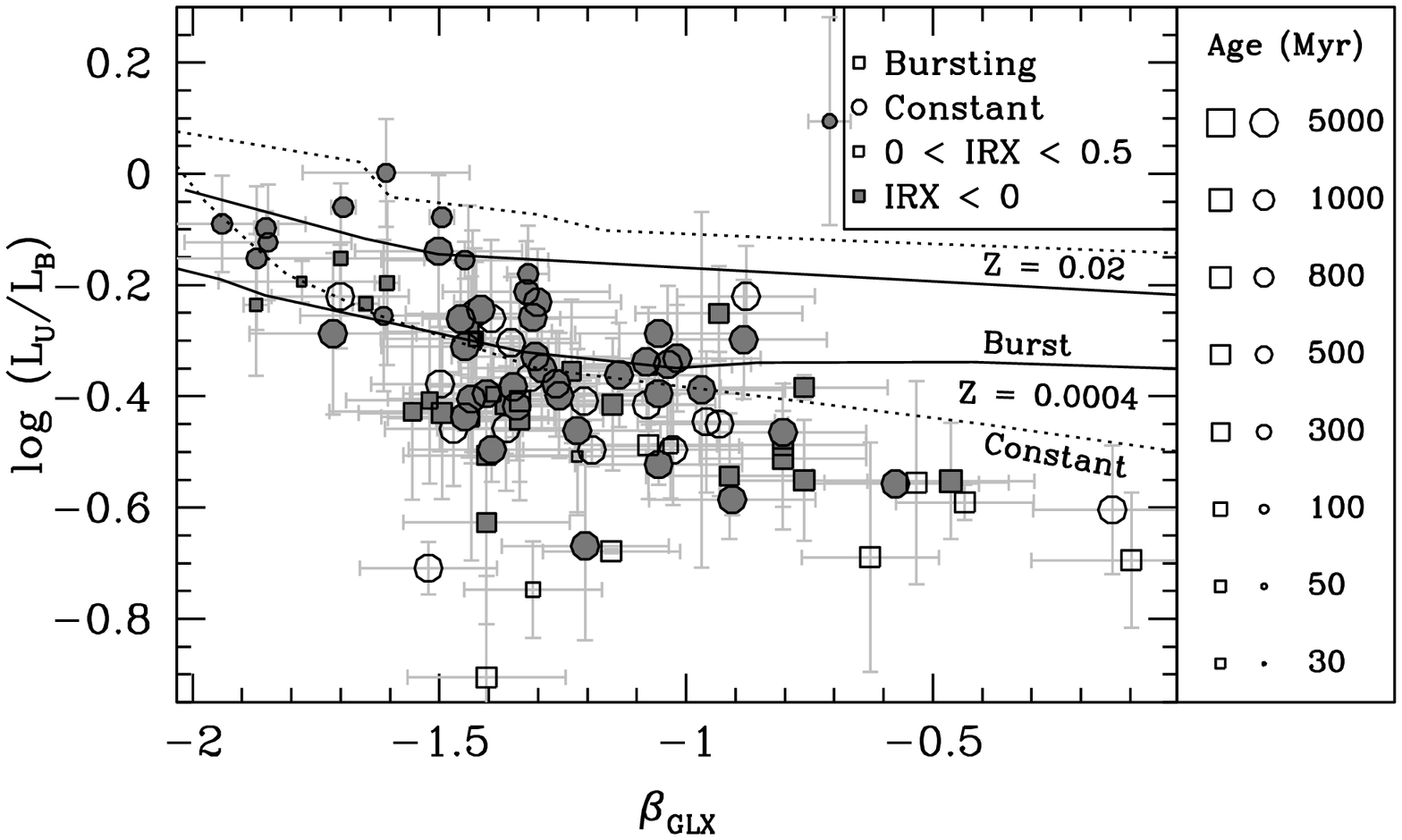}
\plotone{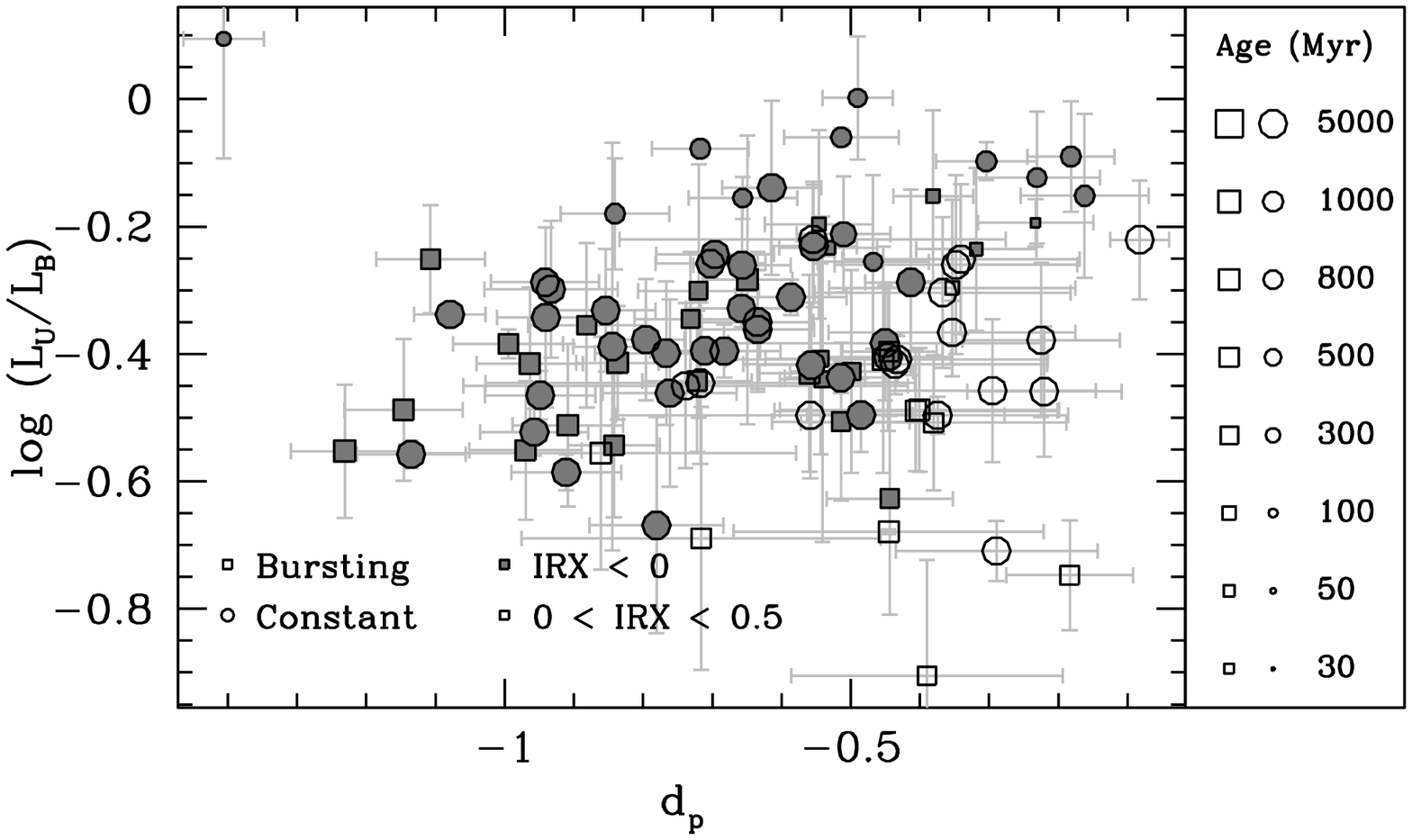}
\caption{$Top$:  The extinction-corrected U$-$B color as a function of the UV color $\beta_{\rm GLX}$.  U$-$B is expressed as the ratio of the fluxes of the two bands, $\log\ ({\rm L}_{U}/{\rm L}_{B})$.  The symbols are the same as in Figure \ref{fig:met}.  The gray bars represent the 1$\sigma$ errors.  The lines are predicted trends for the age as a function of $\beta_{\rm GLX}$ by Starburst99 models for constant star-forming (dotted lines) and bursting (solid lines) systems at metallicities of $Z=0.0004$ and 0.02.  
$Bottom$:  The extinction-corrected U$-$B colors as a function of the perpendicular distance $d_p$.  
\label{fig:UB}}
\end{figure}

The U$-$B color is a sensitive indicator of a stellar population mean age (Figure \ref{fig:UB}).  The two filters straddle the break at 4000 \AA, which is the strongest discontinuity in the UV-to-NIR spectrum of a stellar population, and arises because of the sharp opacity edge produced by the accumulation of a large number of spectral lines in a narrow wavelength region \citep{bruzual83,kauffmann03}.  The discontinuity is small in young, massive stars, because the elements that produce the lines are multiply ionized, but increases as the population ages.  We employ the extinction-corrected U$-$B color, expressed as the ratio of the U-to-B luminosity, as an indicator of the stellar population mean age, and plot it as a function of both $\beta_{\rm GLX}$ and $d_p$ (Figure~\ref{fig:UB}).  Predictions from the Starburst99 models are also shown for the U$-$B versus $\beta_{\rm GLX}$ plot.  Both B-galaxies and C-galaxies have significant correlation with $\beta_{\rm GLX}$ at the 5$\sigma$ level when combined together (Table 5).  The model expectations are in reasonable agreement with the data; the major deviations are observed for the galaxies in the highest IRX bin (IRX $>0$), which also tend to have the largest error bars (Figure~\ref{fig:UB}, top panel).  No significant correlation (both Spearman and Kendall coefficients indicating below 4$\sigma$ significance) is observed between U$-$B and $d_p$.  This, together with the results of section~5.1, leads us to conclude that the mean stellar population age, as traced by U$-$B colors, is not responsible for the perpendicular deviation of normal star-forming galaxies from the starburst attenuation relation. 

\subsection{Birthrate Parameter}
The birthrate parameter, $b$, is the ratio of the current star formation to its overall lifetime average \citep{kennicutt08}.  Following previous studies, we use both the ratio of the FUV to near-IR (NIR) luminosities and the \ewha\ as proxies for the birthrate parameter.  We correct both parameters for the effects of dust attenuation using the best-fit $E(B-V)$ of Figure \ref{fig:E} for each galaxy, instead of employing mean correction factors like other studies \citep{cortese06,dale09}.  It should be remembered that, in bursting galaxies, the FUV-to-NIR ratio and the \ewha\ are better proxies of the mean age of the young populations than of the birthrate parameter as the SED is dominated by the most recent stellar population.

\subsubsection{The FUV to NIR Luminosity Ratio}
The FUV (1520~\AA) traces star formation activity over very recent times ($\sim$100~Myr) while the NIR (3.6~\micron) traces the total stellar mass built up over much longer timescales \citep{dale09}.  The ratio roughly gives the SFR per unit stellar mass, providing a normalized measure of the star formation activity, but is very sensitive to extinction effects, which, as mentioned above, we remove, by correcting the FUV luminosity.  The effect of dust extinction on the 3.6 micron luminosity is negligible.  The FUV/NIR ratio is calculated as: 
\begin{equation}\label{eq:8}
{\rm FUV/NIR} = \frac{\nu L_{\nu,corr}(1520~{\rm \AA})}{\nu L_{\nu}(3.6~\micron)}.
\end{equation}

\begin{figure}
\epsscale{1.2}
\plotone{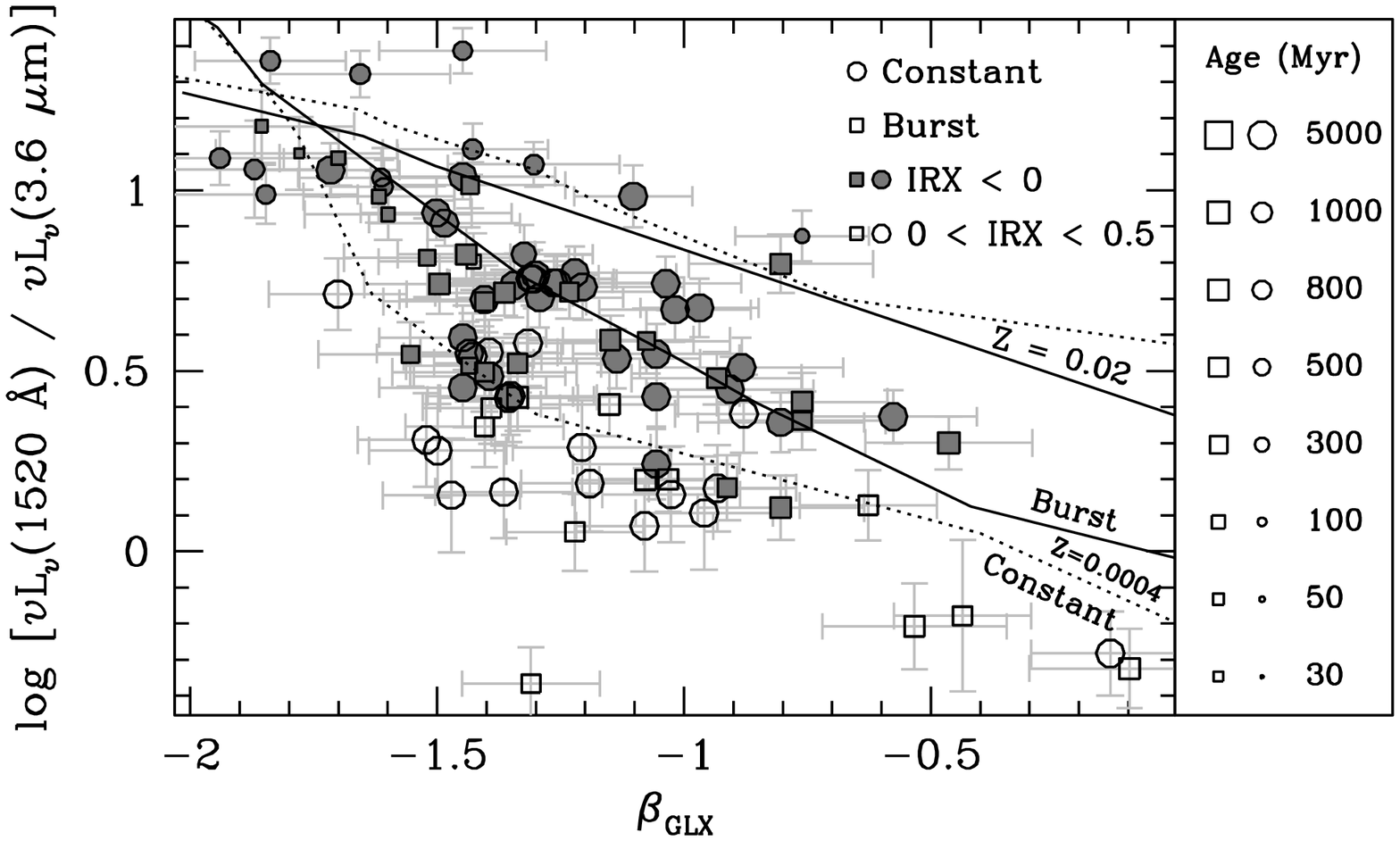}
\plotone{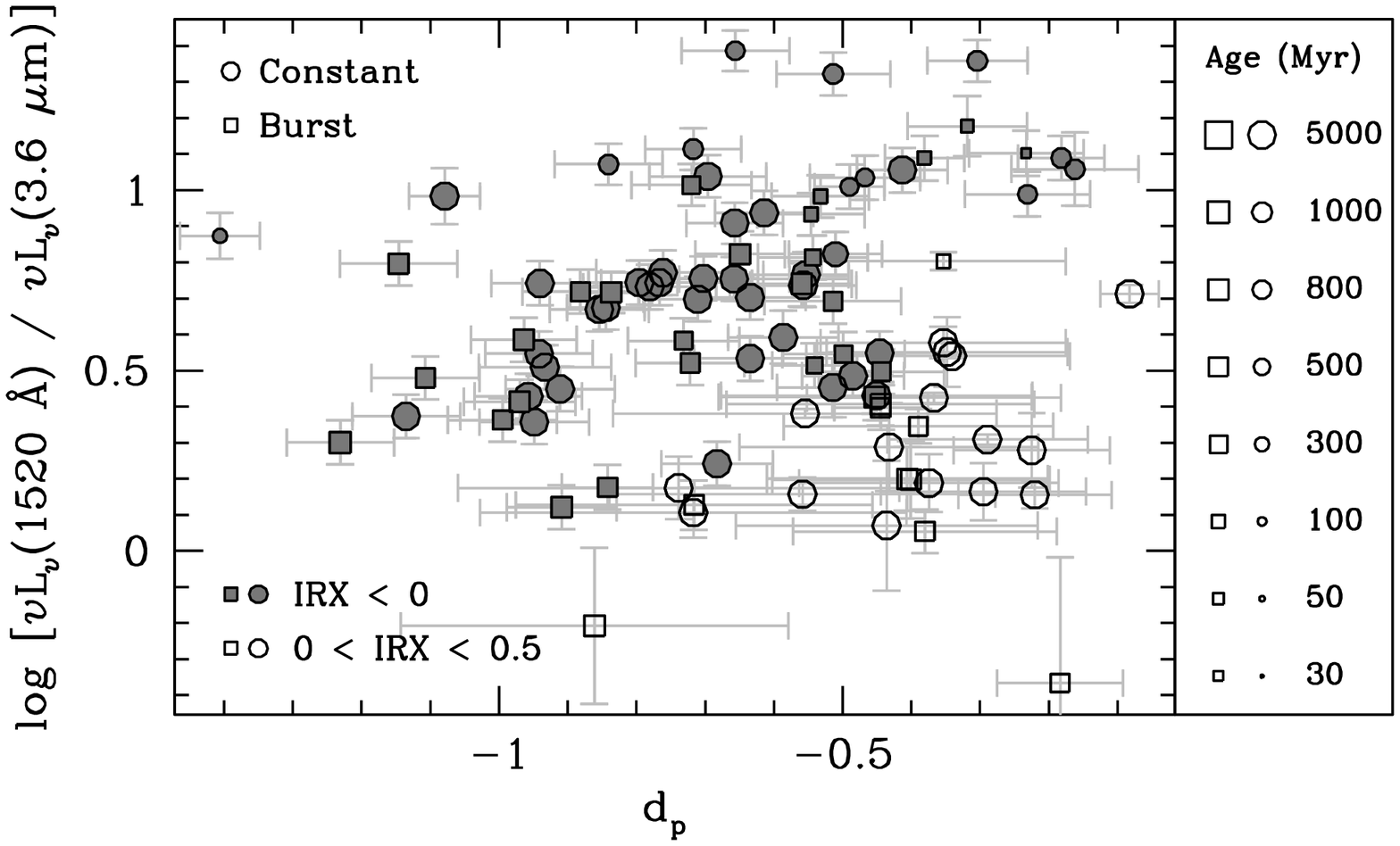}
\caption{$Top$:  The extinction-corrected FUV/NIR ratio of each galaxy as a function of the UV color $\beta_{\rm GLX}$.  The symbol and lines are the same as in Figure \ref{fig:met}.  The gray bars represent the 1$\sigma$ errors.  Model expectations from Starburst99 are shown as in Figure \ref{fig:UB}.
$Bottom$:  The extinction-corrected FUV/NIR ratio of each galaxy as a function of $d_p$, the perpendicular distance to the starburst IRX-$\beta$ relation.  
\label{fig:nL}}
\end{figure}

Figure \ref{fig:nL} shows the extinction-corrected FUV/NIR ratio as a function of $\beta_{\rm GLX}$ and $d_p$ in the top and bottom panel, respectively.  There is considerable spread between the FUV/NIR and $d_p$ for the whole sample and we do not find any correlation between these two quantities.  When we only examine the galaxies at IRX $\leq 0$, there is a correlation with modest significance (4$\sigma$ or below; Table 5) between the FUV/NIR ratio vs $d_p$.  These results are quantitatively consistent with those of \citet{dale09}, when we limit their sample to the IRX $\leq 0.5$ galaxies only. 

Conversely, we find a significant correlation between the FUV/NIR ratio and $\beta_{\rm GLX}$, with values of the correlation coefficients between 5$\sigma$ and 6$\sigma$.  This is the strongest correlation we find in our entire analysis.  The expectations from the Starburst99 models reproduce well the observes trend.  As in previous cases, the low-metallicity models agree with the data better than the high-metallicity models, as one would expect from the fact that our sample is mostly populated by low-metallicity galaxies.

\subsubsection{H$\alpha$ Equivalent Width}
\begin{figure}
\epsscale{1.2}
\plotone{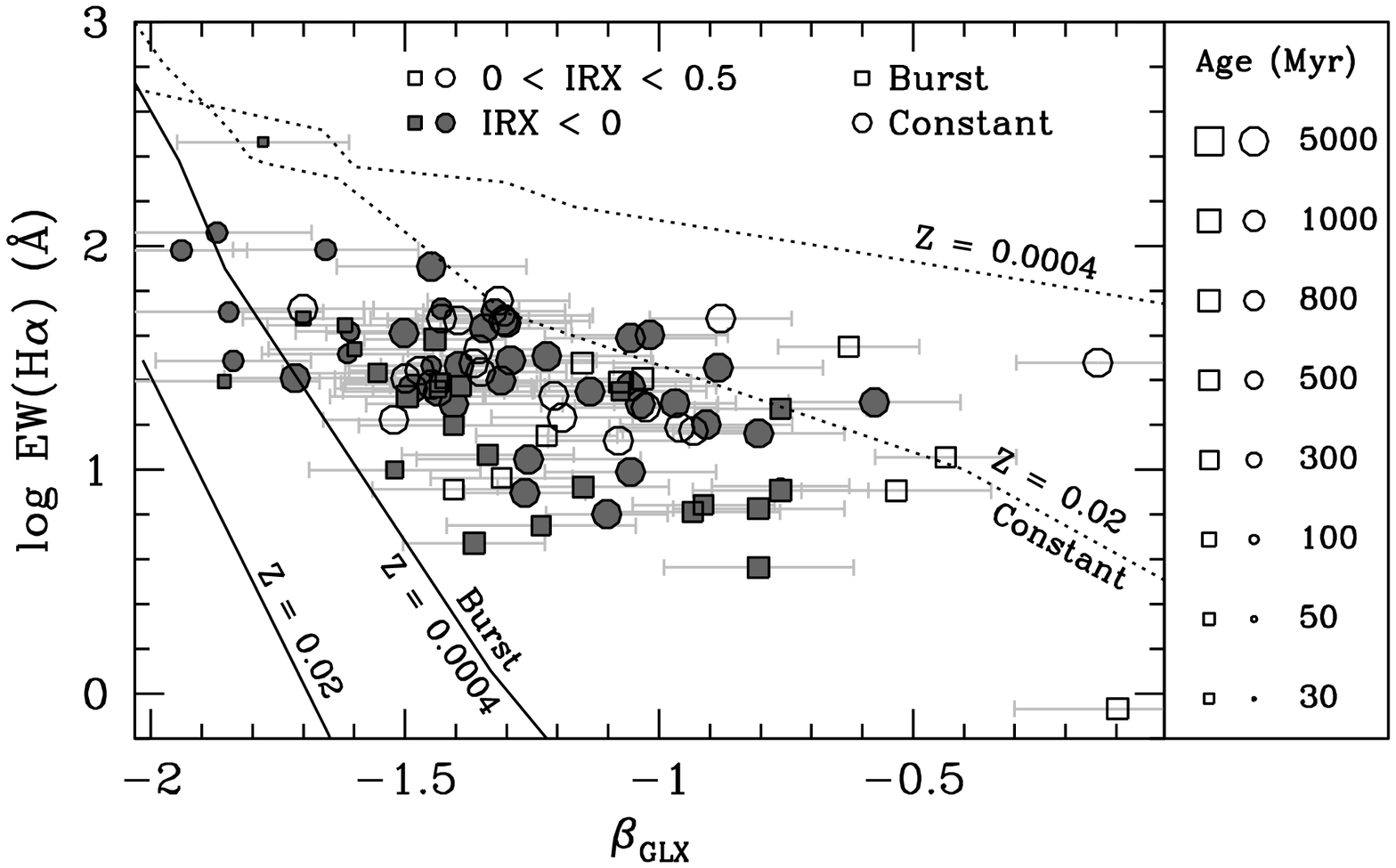}
\plotone{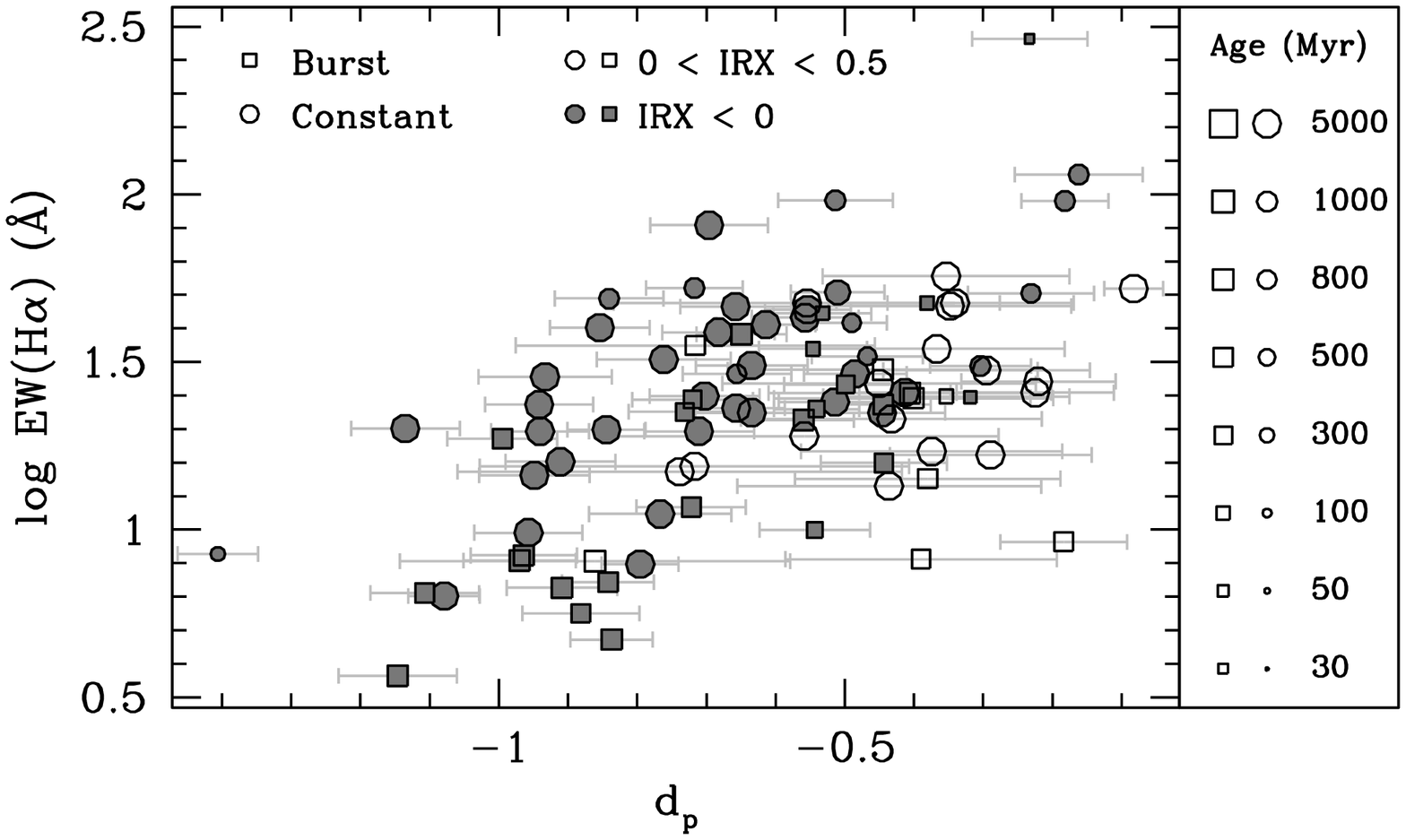}
\caption{$Top$:  The extinction-corrected H$\alpha$ equivalent width, \ewha, of each galaxy as a function of the UV spectral index $\beta_{\rm GLX}$.  The symbol and lines are the same as in Figure \ref{fig:met}.  The gray bars represent the 1$\sigma$ errors.  Model expectations from Starburst99 are shown as in Figure \ref{fig:UB}.  $Bottom$:  The extinction-corrected \ewha\ of each galaxy as a function of the perpendicular distance $d_p$.  
\label{fig:ewha2}}
\end{figure}

The \ewha\ is less sensitive than the FUV/NIR ratio to the effects of dust attenuation, but still needs to be corrected for it, because of the differential attenuation affecting the line emission (originating from HII regions surrounding the youngest stars, and, thus, sensitive to timescales $<$10~Myr) and the underlying continuum \citep[originating from older stars over much longer timescales;][]{kennicutt94}.  The differential attenuation between line and continuum has been reported by multiple authors \citep[e.g.,][]{calzetti94,hao11} and we implement it as: 
\begin{equation}\label{eq:9}
\rm{EW}(\rm{H}\alpha)_{\rm corr} = EW(\rm{H}\alpha)\cdot 10^{+0.4\ E(B-V)[\frac{2.535}{0.44}-k(\rm{H}\alpha)]},
\end{equation}
where k(H$\alpha$) is the attenuation measured in the continuum, using the attenuation curve of \citet{calzetti00}.  We take the \ewha\ for our sample from the measurements of \citep{kennicutt08}, which we correct for [NII] contamination using the [NII] measurements from the same paper.  We obtain \ewha$_{\rm corr}$ values for 93 out of 98 galaxies, reported in Figure \ref{fig:ewha2} as a function of both $\beta_{\rm GLX}$ and $d_p$.  For the plot as a function of $\beta_{\rm GLX}$, expectations from the Starburst99 models are also shown. 

Both trends are significant at the $\sim$5$\sigma$ level (Table 5) for the whole sample, implying that the \ewha\ decreases both for larger $\beta_{\rm GLX}$ (redder UV colors) and larger distance $d_p$ from the starburst relation.  The models bracket the observed values, although only the constant star formation model at low-metallicity goes through the data points.  This lack of overlap may not be surprising, as the H$\alpha$ emission is extremely sensitive to the most recent star formation, while our SED modeling (which did not include line emission) is sensitive only to star formation timescales of 100~Myr (FUV) or longer.  Thus, even a modest burst of star formation would increase the \ewha, without necessarily affecting the UV colors, especially if the burst occurs in a dust-obscured region \citep{calzetti94,zaritsky04}.  This is a limitation of our study; by modeling only the continuum SED, we do not capture the exact timescale of the star formation.  This limitation is compounded by the fact that we do not have sufficient information to include, in our modeling, variations of dust attenuation across each galaxy.

\section{Discussion}\label{sec:6}
Dust-poor star-forming galaxies offer a unique opportunity to explore the impact of stellar population age on the spread in the IRX-$\beta$ relation, by mitigating or eliminating the need for uncertain assumptions on the dust extinction curve and geometry.

When the FUV-to-NIR SED of galaxies can be modeled by a burst of star formation, we find significant correlations (significance of 4$\sigma$ or higher as evaluated from the Spearman and Kendall coefficients) between stellar population mean age estimators and $\beta_{\rm GLX}$, i.e., the measured UV color.  Specifically, $\beta_{\rm GLX}$ correlates with the mean age directly obtained from the SED fitting, and with proxies for the mean age:  the FUV/NIR ratio and the U$-$B color.  For galaxies whose FUV-to-NIR SED is best approximated with constant star formation, $\beta_{\rm GLX}$ is best correlated with the FUV/NIR ratio and the \ewha, which here represents a proxy for the birthrate parameter.  The combined sample also show the strongest correlations (significance of 5$\sigma$ or higher) with age/duration proxies:  U$-$B color, FUV/NIR ratio, and \ewha.  In most cases, the correlation with the perpendicular distance $d_p$ is not significant, or is considerably less significant than the correlation with $\beta_{\rm GLX}$, except for the case in which the dependent variable is the \ewha.  In summary, our data allow us to conclude that: (1) variations in the mean age of the stellar population can account for the spread in the IRX-$\beta$ relation in low-dust-content galaxies; and (2) that the spread is better correlated with $\beta_{\rm GLX}$ than with $d_p$. 

While a dependency of the spread on the mean age, and/or birthrate parameter, and/or SFH has been already suggested \citep[e.g.,][]{kong04,dale09,munoz09,boquien12}, this dependency has always been suggested to be a function of the perpendicular distance (i.e., shortest distance) from the starburst relation.  Our results show that the main independent variable is the observed UV color, as measured by $\beta_{\rm GLX}$.  Our analyses do not support earlier results that disfavor the role of the age/birthrate of the stellar population \citep[e.g.,][]{seibert05,cortese06,johnson07}.  We suspect these results may have been affected by the degeneracy between age and dust attenuation, as they have analyzed galaxies with a large range of dust content.

A majority of the galaxies in the LVL survey shows evidence that the outer edges were older than the inner regions of the galaxies \citep{dale09}.  Our simplistic treatment of the SFH of our sample galaxies is unlikely to have captured these complexities, beyond attributing a luminosity-weighted age to the observed SEDs.  However, the old central stellar populations in the LVL galaxies may account for the observed spread in the IRX-$\beta$ relation, and explain their deviations from the starburst attenuation curve.  Indeed, many studies that focus on the galactocentric behavior of the IRX-$\beta$ relation find a trend between the distance of a region from the galaxy center and its location on the IRX-$\beta$ diagram \citep{gordon04,calzetti05,perez06,thilker07,munoz09,boquien09,boquien12,mao12}.

Our metal-poor galaxies do not display a correlation between either metallicity and $\beta_{\rm GLX}$ or metallicity and $d_p$, independently of their SFH (bursting or constant).  This means that galaxies with redder UV spectral colors do not necessarily represent galaxies with higher metallicities.  This is different from what found by \citet{cortese06}, where a correlation, at the 3$\sigma$ level, was observed between metallicity and $d_p$.  Our strongest correlation is less than 2$\sigma$.  We interpret this discrepancy as an effect of the primary correlation between IRX and $\beta_{\rm GLX}$ (the starburst relation) in the Cortese et al.'s data, rather than an effect of the secondary correlation (that drives the spread around the primary relation).  In fact, the sample analyzed by \citet{cortese06} contains mostly dust-rich galaxies.

\section{Summary and Conclusions}\label{sec:7}
We have modeled the multi-wavelength, FUV-to-FIR, SEDs of 98 nearby ($\le$11~Mpc) star-forming galaxies selected from the $Spitzer$ LVL survey to contain little or no dust on the IRX-$\beta$ diagram (IRX $=\log L_{\rm TIR}/L_{\rm FUV}\le 0.5$).  On account of this selection, our sample contains mostly low-metallicity dwarf and irregular galaxies, although it still displays a significant spread in $\beta$.  With an average SFR$_{\rm FUV}$ of 0.04~M$_{\sun}$ yr$^{-1}$ \citep{lee09a,lee09b} and a mean SFR/area $\sim$5$\times10^{-4}$ M$_{\sun}$~yr$^{-1}$ kpc$^{-2}$ \citep{calzetti10}, our galaxies are representative of local, normal star-forming systems.

Synthetic model spectra are generated from Starburst99, with two simple star formation histories (instantaneous bursts and constant star formation) in the age range 10~Myr to 5~Gyr, a Kroupa IMF, and a range of metallicity values from $Z = 0.0004$ to 0.020.  For the 2/3 of the sample with measured oxygen abundances, we match the models to the measured metallicity.  Although our systems tend to be dust-poor, we still apply dust attenuation to the model SEDs, using the starburst curve of \citet{calzetti00}.

We then plot the model-fitted ages (durations, for the constant star formation models) and other estimators of age/duration, such as FUV/NIR ratio, the U$-$B color, and the \ewha\ against both the measured UV colors ($\beta_{\rm GLX}$) and the perpendicular distance $d_p$ from the starburst IRX-$\beta$ relation of \citet{overzier11}.  In order to avoid any impact from dust attenuation, we plot the extinction-corrected values of the age/duration estimators, using the color excess values returned by the best-fit model SEDs.  The observed trends between age/duration estimators and UV color are also compared against direct predictions from the Starburst99 models.  Finally, we explore the color excess and oxygen abundance as a function of both $\beta_{\rm GLX}$ and $d_p$.

Our main results can be summarized as follows:  
\begin{enumerate}
\item The galaxies show a clear trend between age estimators (extinction-corrected FUV/NIR, \ewha, and U$-$B) and the UV color, $\beta_{\rm GLX}$.  These trends are mostly driven by the galaxies that are best-fit with instantaneous bursts (as opposed to constant star-formation);
\item The instantaneous burst galaxies also show a correlation between UV colors and the age directly determined from the best-fit SED model; 
\item Conversely, we usually find non-significant correlations between any parameter and the perpendicular distance $d_p$, with the only exception of the \ewha, which shows a significance comparable to that of \ewha\ and $\beta_{\rm GLX}$;
\item Color excess and oxygen abundance do not correlate with either $\beta_{\rm GLX}$ or $d_p$.
\end{enumerate}

We conclude that the ``second parameter'' for the IRX-$\beta$ relation is most likely the mean population age, rather than variations in the dust extinction curve and/or geometry.  This has already been concluded by many other authors (see list in the Introduction), although they usually advocate the correlation to be with the perpendicular distance $d_p$.  We find, instead, that the mean age correlates most strongly and more directly with the measured UV color.  We believe that the earlier findings may have been affected by effects of age/extinction degeneracy, as the samples usually include both dusty and dust-poor systems.  As our analysis contains only 98 galaxies, a larger, more statistically significant sample may be required to confirm our results.

\acknowledgements
We thank the anonymous referee for the careful and constructive comments that greatly improved this paper.  
This research has made use of the NASA/IPAC Extragalactic Database (NED) which is operated by the Jet Propulsion Laboratory, California Institute of Technology, under contract with NASA.  
Funding for SDSS and SDSS-II has been provided by the Alfred P. Sloan Foundation, the Participating Institutions, the National Science Foundation, the U.S.  Department of Energy, the National Aeronautics and Space Administration, the Japanese Monbukagakusho, the Max Planck Society, and the Higher Education Funding Council for England.  The SDSS Web Site is http://www.sdss.org/.  The SDSS is managed by the Astrophysical Research Consortium for the Participating Institutions.  
This publication makes use of data products from the Two Micron All Sky Survey, which is a joint project of the University of Massachusetts and the Infrared Processing and Analysis Center/California Institute of Technology, funded by the National Aeronautics and Space Administration and the National Science Foundation.
We gratefully acknowledge NASAs support for the GALEX mission, developed in cooperation with the Centre National dEtudes Spatiales of France and the Korean Ministry of Science and Technology.  \\

\begin{deluxetable}{lccccccccc}
\tabletypesize{\scriptsize}
\tablecaption{Probabilities for Bursting Star-Forming Galaxies \label{tab:prob}}
\tablecolumns{10}
\tablewidth{0pt}
\tablehead{
\colhead{} &
\multicolumn{3}{c}{Total} & 
\multicolumn{3}{c}{IRX $ \leq 0$} & 
\multicolumn{3}{c}{$0<$ IRX $ \leq 0.5$} 
\\ \cline{2-10} \noalign{\smallskip}
\colhead{Variables} &
\colhead{$N$} &
\colhead{Spearman} &
\colhead{Kendall} & 
\colhead{$N$} &
\colhead{Spearman} &
\colhead{Kendall} & 
\colhead{$N$} &
\colhead{Spearman} &
\colhead{Kendall} 
}
\startdata
Age vs $\beta_{\rm GLX}$ & 38 & $\rho=0.69$ & $\tau=0.57$ & 25 & $\rho=0.76$ & $\tau=0.65$ & 13 & $\rho=0.58$ & $\tau=0.48$ \\
            & & 4.2$\sigma$ & {\bf 5.1$\sigma$} & & 3.7$\sigma$ & {\bf 4.6$\sigma$} & & 2.0$\sigma$ & 2.3$\sigma$\\
Age vs $d_p$ & 36 & $\rho=-0.47$ & $\tau=-0.38$ & 25 & $\rho=-0.77$ & $\tau=-0.61$ & 11 & $\rho=-0.67$ & $\tau=-0.52$\\
            & & 2.8$\sigma$ & 3.2$\sigma$ & & 3.8$\sigma$ & 4.3$\sigma$ & & 2.1$\sigma$ & 2.3$\sigma$\\
$E(B-V)$ vs $\beta_{\rm GLX}$ & 38 & $\rho=0.23$ & $\tau=0.15$ & 25 & $\rho=-0.013$ & $\tau=-0.03$ & 13 & $\rho=0.34$ & $\tau=0.25$ \\
            & & 1.4$\sigma$ & 1.3$\sigma$ & & 0.06$\sigma$ & 0.18$\sigma$ & & 1.1$\sigma$ & 1.2$\sigma$\\
$E(B-V)$ vs $d_p$ & 36 & $\rho=0.39$ & $\tau=0.30$ & 25 & $\rho=0.31$ & $\tau=0.24$ & 11 & $\rho=0$ & $\tau=0$ \\
            & & 2.3$\sigma$ & 2.5$\sigma$ & & 1.5$\sigma$ & 1.7$\sigma$ & & 0$\sigma$ & 0$\sigma$\\
12+log(O/H) vs $\beta_{\rm GLX}$ & 24 & $\rho=0.41$ & $\tau=0.31$ & 14 & $\rho=0$ & $\tau=0$ & 10 & $\rho=-0.58$ & $\tau=-0.36$ \\
            & & 2.0$\sigma$ & 2.2$\sigma$ & & 0$\sigma$ & 0$\sigma$ & & 1.8$\sigma$ & 1.4$\sigma$\\
12+log(O/H) vs $d_p$ & 22 & $\rho=0.36$ & $\tau=0.29$ & 14 & $\rho=0.13$ & $\tau=0.03$ & 8 & $\rho=0.70$ & $\tau=0.55$ \\
            & & 1.7$\sigma$ & 1.9$\sigma$ & & 0.5$\sigma$ & 0.2$\sigma$ & & 1.8$\sigma$ & 1.9$\sigma$\\
U$-$B vs $\beta_{\rm GLX}$ & 38 & $\rho=-0.64$ & $\tau=-0.49$ & 25 & $\rho=-0.68$ & $\tau=-0.51$ & 13 & $\rho=-0.28$ & $\tau=-0.27$ \\
            & & 3.9$\sigma$ & 4.3$\sigma$ & & 3.3$\sigma$ & 3.6$\sigma$ & & 1.0$\sigma$ & 1.2$\sigma$\\
U$-$B vs $d_p$ & 36 & $\rho=0.07$ & $\tau=0.10$ & 25 & $\rho=0.44$ & $\tau=0.34$ & 11 & $\rho=0$ & $\tau=0.02$ \\
            & & 0.4$\sigma$ & 0.8$\sigma$ & & 2.2$\sigma$ & 2.4$\sigma$ & & 0$\sigma$ & 0$\sigma$\\
FUV/NIR vs $\beta_{\rm GLX}$ & 38 & $\rho=-0.78$ & $\tau=-0.61$ & 25 & $\rho=-0.80$ & $\tau=-0.65$ & 13 & $\rho=-0.66$ & $\tau=-0.51$ \\
            & & {\bf 4.7$\sigma$} & {\bf 5.4$\sigma$} & & 3.9$\sigma$ & {\bf 4.5$\sigma$} & & 2.3$\sigma$ & 2.4$\sigma$\\ 
FUV/NIR vs $d_p$ & 36 & $\rho=0.09$ & $\tau=0.04$ & 25 & $\rho=0.59$ & $\tau=0.41$ & 11 & $\rho=-0.05$ & $\tau=-0.09$ \\
            & & 0.5$\sigma$ & 0.4$\sigma$ & & 2.9$\sigma$ & 2.9$\sigma$ & & 0.14$\sigma$ & 0.4$\sigma$\\ 
\ewha\ vs $\beta_{\rm GLX}$ & 35 & $\rho=-0.58$ & $\tau=-0.43$ & 24 & $\rho=-0.78$ & $\tau=-0.58$ & 11 & $\rho=-0.22$ & $\tau=-0.09$ \\
            & & 3.4$\sigma$ & 3.6$\sigma$ & & 3.8$\sigma$ & 4.0$\sigma$ & & 0.7$\sigma$ & 0.4$\sigma$\\ 
\ewha\ vs $d_p$ & 33 & $\rho=0.60$ & $\tau=0.40$ & 24 & $\rho=0.73$ & $\tau=0.57$ & 9 & $\rho=-0.32$ & $\tau=-0.39$ \\
            & & 3.3$\sigma$ & 3.3$\sigma$ & & 3.8$\sigma$ & 3.7$\sigma$ & & 1.0$\sigma$ & 1.5$\sigma$
\enddata
\tablecomments{
The Spearman ($\rho$) and Kendall ($\tau$) rank correlation coefficients for the variables shown in the first column for (1) the entire sample of bursting galaxies, (2) the bursting galaxies in the range IRX $= \log L_{\rm TIR}/L_{\rm FUV}\leq0$, and (3) the galaxies in the range $0 < \log L_{\rm TIR}/L_{\rm FUV}\leq0.5$, where $N$ represents the number of galaxies in each sample.  Below the correlation coefficients are the significance of the correlation assuming Gaussian statistics.  Values that are bold have a significance of 4.5$\sigma$ or greater.  
} 
\end{deluxetable}

\begin{deluxetable}{lccccccccc}
\tabletypesize{\scriptsize}
\tablecaption{Probabilities for Constant Star-Forming Galaxies \label{tab:probc}}
\tablecolumns{10}
\tablewidth{0pt}
\tablehead{
\colhead{} &
\multicolumn{3}{c}{Total} & 
\multicolumn{3}{c}{IRX $ \leq 0$} & 
\multicolumn{3}{c}{$0<$ IRX $ \leq 0.5$} 
\\ \cline{2-10} \noalign{\smallskip}
\colhead{Variables} &
\colhead{$N$} &
\colhead{Spearman} &
\colhead{Kendall} & 
\colhead{$N$} &
\colhead{Spearman} &
\colhead{Kendall} & 
\colhead{$N$} &
\colhead{Spearman} &
\colhead{Kendall} 
}
\startdata
Age vs $\beta_{\rm GLX}$ & 60 & $\rho=0.41$ & $\tau=0.33$ & 43 & $\rho=0.46$ & $\tau=0.37$ & 17 & $\rho=0$ & $\tau=0$ \\
            & & 3.1$\sigma$ & 3.7$\sigma$ & & 3.0$\sigma$ & 3.5$\sigma$ & & 0$\sigma$ & 0$\sigma$\\
Age vs $d_p$ & 59 & $\rho=-0.15$ & $\tau=-0.13$ & 43 & $\rho=-0.35$ & $\tau=-0.28$ & 16 & $\rho=0$ & $\tau=0$ \\
            & & 1.2$\sigma$ & 1.4$\sigma$ & & 2.3$\sigma$ & 2.7$\sigma$ & & 0$\sigma$ & 0$\sigma$\\
$E(B-V)$ vs $\beta_{\rm GLX}$ & 60 & $\rho=0.21$ & $\tau=0.15$ & 43 & $\rho=0.26$ & $\tau=0.19$ & 17 & $\rho=-0.010$ & $\tau=0$ \\
            & & 1.6$\sigma$ & 1.7$\sigma$ & & 1.7$\sigma$ & 1.8$\sigma$ & & 0.04$\sigma$ & 0$\sigma$\\
$E(B-V)$ vs $d_p$ & 59 & $\rho=0.40$ & $\tau=0.31$ & 43 & $\rho=0.13$ & $\tau=0.09$ & 16 & $\rho=0.06$ & $\tau=0.04$\\
            & & 3.1$\sigma$ & 3.5$\sigma$ & & 0.8$\sigma$ & 0.9$\sigma$ & & 0.2$\sigma$ & 0.2$\sigma$ \\
12+log(O/H) vs $\beta_{\rm GLX}$ & 36 & $\rho=0.44$ & $\tau=0.31$ & 27 & $\rho=0.46$ & $\tau=0.33$ & 9 & $\rho=0.20$ & $\tau=0.06$ \\
            & & 2.6$\sigma$ & 2.7$\sigma$ & & 2.4$\sigma$ & 2.4$\sigma$ & & 0.6$\sigma$ & 0.2$\sigma$\\
12+log(O/H) vs $d_p$ & 35 & $\rho=0.10$ & $\tau=0.07$ & 27 & $\rho=-0.14$ & $\tau=-0.12$ & 8 & $\rho=-0.19$ & $\tau=-0.14$ \\
            & & 0.6$\sigma$ & 0.6$\sigma$ & & 0.7$\sigma$ & 0.9$\sigma$ & & 0.5$\sigma$ & 0.5$\sigma$\\
U$-$B vs $\beta_{\rm GLX}$ & 60 & $\rho=-0.49$ & $\tau=-0.36$ & 43 & $\rho=-0.56$ & $\tau=-0.41$ & 17 & $\rho=-0.25$ & $\tau=-0.26$ \\
            & & 3.7$\sigma$ & 4.1$\sigma$ & & 3.6$\sigma$ & 3.9$\sigma$ & & 1.0$\sigma$ & 1.4$\sigma$\\
U$-$B vs $d_p$ & 59 & $\rho=0.22$ & $\tau=0.19$ & 43 & $\rho=0.35$ & $\tau=0.28$ & 16 & $\rho=0.20$ & $\tau=0.18$ \\
            & & 1.9$\sigma$ & 2.1$\sigma$ & & 2.4$\sigma$ & 2.6$\sigma$ & & 0.8$\sigma$ & 0.9$\sigma$\\
FUV/NIR vs $\beta_{\rm GLX}$ & 60 & $\rho=-0.52$ & $\tau=-0.37$ & 43 & $\rho=-0.65$ & $\tau=-0.47$ & 17 & $\rho=-0.52$ & $\tau=-0.38$ \\
            & & 4.0$\sigma$ & 4.2$\sigma$ & & 4.2$\sigma$ & {\bf 4.5$\sigma$} & & 2.1$\sigma$ & 2.1$\sigma$\\ 
FUV/NIR vs $d_p$ & 59 & $\rho=-0.05$ & $\tau=-0.02$ & 43 & $\rho=0.38$ & $\tau=0.28$ & 16 & $\rho=0.40$ & $\tau=0.27$ \\
            & & 0.3$\sigma$ & 0.2$\sigma$ & & 2.5$\sigma$ & 2.7$\sigma$ & & 1.5$\sigma$ & 1.4$\sigma$\\ 
\ewha\ vs $\beta_{\rm GLX}$ & 58 & $\rho=-0.52$ & $\tau=-0.36$ & 40 & $\rho=-0.60$ & $\tau=-0.43$ & 18 & $\rho=-0.29$ & $\tau=-0.24$ \\
            & & 3.9$\sigma$ & 4.0$\sigma$ & & 3.8$\sigma$ & 3.9$\sigma$ & & 1.2$\sigma$ & 1.4$\sigma$\\ 
\ewha\ vs $d_p$ & 57 & $\rho=0.47$ & $\tau=0.37$ & 40 & $\rho=0.61$ & $\tau=0.44$ & 17 & $\rho=0.44$ & $\tau=0.34$ \\
            & & 3.7$\sigma$ & 4.1$\sigma$ & & 3.8$\sigma$ & 4.0$\sigma$ & & 1.8$\sigma$ & 1.9$\sigma$   
\enddata
\tablecomments{
The Spearman ($\rho$) and Kendall ($\tau$) rank correlation coefficients for the variables listed in the first column for (1) the entire sample of constant star-forming galaxies, (2) the constant star-forming galaxies in the range $\log L_{\rm TIR}/L_{\rm FUV}\leq 0$, and (3) the galaxies in the range $0 < \log L_{\rm TIR}/L_{\rm FUV} \leq 0.5$, where $N$ represents the number of galaxies in each sample.  Below the correlation coefficients are the significance of the correlation assuming Gaussian statistics.  Values that are bold have a significance of 4.5$\sigma$ or greater.  
} 
\end{deluxetable}

\begin{deluxetable}{lccccccccc}
\tabletypesize{\scriptsize}
\tablecaption{Probabilities for Entire Sample of Star-Forming Galaxies \label{tab:tot}}
\tablecolumns{10}
\tablewidth{0pt}
\tablehead{
\colhead{} &
\multicolumn{3}{c}{Total} & 
\multicolumn{3}{c}{IRX $ \leq 0$} & 
\multicolumn{3}{c}{$0<$ IRX $ \leq 0.5$} 
\\ \cline{2-10} \noalign{\smallskip}
\colhead{Variables} &
\colhead{$N$} &
\colhead{Spearman} &
\colhead{Kendall} & 
\colhead{$N$} &
\colhead{Spearman} &
\colhead{Kendall} & 
\colhead{$N$} &
\colhead{Spearman} &
\colhead{Kendall} 
}
\startdata
Age vs $\beta_{\rm GLX}$ & 98 & $\rho=0.15$ & $\tau=0.12$ & 68 & $\rho=0.27$ & $\tau=0.21$ & 30 & $\rho=-0.13$ & $\tau=-0.09$ \\
            & & 1.5$\sigma$ & 1.7$\sigma$ & & 2.2$\sigma$ & 2.5$\sigma$ & & 0.7$\sigma$ & 0.7$\sigma$\\
Age vs $d_p$ & 95 & $\rho=-0.07$ & $\tau=-0.05$ & 68 & $\rho=-0.18$ & $\tau=-0.13$ & 27 & $\rho=0.15$ & $\tau=0.10$ \\
            & & 0.6$\sigma$ & 0.7$\sigma$ & & 1.5$\sigma$ & 1.5$\sigma$ & & 0.7$\sigma$ & 0.7$\sigma$\\
$E(B-V)$ vs $\beta_{\rm GLX}$ & 98 & $\rho=0.22$ & $\tau=0.15$ & 68 & $\rho=0.15$ & $\tau=0.10$ & 30 & $\rho=-0.05$ & $\tau=-0.03$ \\
            & & 2.1$\sigma$ & 2.2$\sigma$ & & 1.3$\sigma$ & 1.2$\sigma$ & & 0.2$\sigma$ & 0.2$\sigma$\\
$E(B-V)$ vs $d_p$ & 95 & $\rho=0.40$ & $\tau=0.46$ & 68 & $\rho=0.19$ & $\tau=0.15$ & 27 & $\rho=0.11$ & $\tau=0.09$ \\
            & & 3.7$\sigma$ & {\bf 4.6$\sigma$} & & 1.6$\sigma$ & 2.1$\sigma$ & & 0.6$\sigma$ & 0.6$\sigma$\\
12+log(O/H) vs $\beta_{\rm GLX}$ & 60 & $\rho=0.31$ & $\tau=0.21$ & 41 & $\rho=0.30$ & $\tau=0.22$ & 19 & $\rho=-0.20$ & $\tau=-0.12$ \\
            & & 2.3$\sigma$ & 2.4$\sigma$ & & 1.9$\sigma$ & 2.0$\sigma$ & & 0.9$\sigma$ & 0.7$\sigma$\\
12+log(O/H) vs $d_p$ & 57 & $\rho=0.18$ & $\tau=0.15$ & 41 & $\rho=-0.07$ & $\tau=-0.05$ & 16 & $\rho=0.18$ & $\tau=0.12$ \\
            & & 1.3$\sigma$ & 1.6$\sigma$ & & 0.4$\sigma$ & 0.5$\sigma$ & & 0.7$\sigma$ & 0.6$\sigma$\\
U$-$B vs $\beta_{\rm GLX}$ & 98 & $\rho=-0.53$ & $\tau=-0.39$ & 68 & $\rho=-0.55$ & $\tau=-0.40$ & 30 & $\rho=-0.41$ & $\tau=-0.33$ \\
            & & {\bf 5.3$\sigma$} & {\bf 5.7$\sigma$} & & {\bf 4.5$\sigma$} & {\bf 4.8$\sigma$} & & 2.2$\sigma$ & 2.5$\sigma$\\
U$-$B vs $d_p$ & 95 & $\rho=0.21$ & $\tau=0.16$ & 68 & $\rho=0.41$ & $\tau=0.29$ & 27 & $\rho=0.24$ & $\tau=0.19$ \\
            & & 2.2$\sigma$ & 2.3$\sigma$ & & 3.4$\sigma$ & 3.5$\sigma$ & & 1.2$\sigma$ & 1.4$\sigma$\\
FUV/NIR vs $\beta_{\rm GLX}$ & 98 & $\rho=-0.62$ & $\tau=-0.45$ & 68 & $\rho=-0.69$ & $\tau=-0.51$ & 30 & $\rho=-0.64$ & $\tau=-0.46$ \\
            & & {\bf 6.1$\sigma$} & {\bf 6.6$\sigma$} & & {\bf 5.6$\sigma$} & {\bf 6.1$\sigma$} & & 3.4$\sigma$ & 3.6$\sigma$\\
FUV/NIR vs $d_p$ & 95 & $\rho=0.02$ & $\tau=0.016$ & 68 & $\rho=0.47$ & $\tau=0.33$ & 27 & $\rho=0.22$ & $\tau=0.17$ \\
            & & 0.2$\sigma$ & 0.2$\sigma$ & & 3.8$\sigma$ & 4.0$\sigma$ & & 1.4$\sigma$ & 1.3$\sigma$\\
\ewha\ vs $\beta_{\rm GLX}$ & 93 & $\rho=-0.52$ & $\tau=-0.36$ & 64 & $\rho=-0.60$ & $\tau=-0.42$ & 29 & $\rho=-0.33$ & $\tau=-0.24$ \\
            & & {\bf 5.0$\sigma$} & {\bf 5.2$\sigma$} & & {\bf 4.8$\sigma$} & {\bf 4.9$\sigma$} & & 1.8$\sigma$ & 1.8$\sigma$\\
\ewha\ vs $d_p$ & 90 & $\rho=0.49$ & $\tau=0.37$ & 64 & $\rho=0.64$ & $\tau=0.46$ & 26 & $\rho=0.29$ & $\tau=0.22$ \\
            & & {\bf 4.8$\sigma$} & {\bf 5.1$\sigma$} & & {\bf 5.1$\sigma$} & {\bf 5.3$\sigma$} & & 1.5$\sigma$ & 1.6$\sigma$
\enddata
\tablecomments{
The Spearman ($\rho$) and Kendall ($\tau$) rank correlation coefficients for the variables listed in the first column for (1) the entire sample of star-forming galaxies (constant and bursting), (2) the galaxies in the range $\log L_{\rm TIR}/L_{\rm FUV}\leq 0$, and (3) the galaxies in the range $0 < \log L_{\rm TIR}/L_{\rm FUV} \leq 0.5$, where $N$ represents the number of galaxies in each sample.  Below the correlation coefficients are the significance of the correlation assuming Gaussian statistics.  Values that are bold have a significance of 4.5$\sigma$ or greater.  
} 
\end{deluxetable}


\begin{thebibliography}{}
{\footnotesize
\bibitem[Abazajian \etal(2009)]{abazajian09} Abazajian, K.N., \etal\ 2009, \apjs, 182, 543
\bibitem[Ag\"uero \& Paolantonio(1997)]{aguero97} Ag\"uero, E.L. \& Paolantonio, S. 1997, \aj, 114, 102
\bibitem[Bell \etal(2002)]{bell02} Bell, E.F., \etal\ 2002, \apj, 565, 994 
\bibitem[Berg \etal(2012)]{berg12} Berg, D.A., \etal\ 2012, \apj, 754, 98
\bibitem[Boquien \etal(2009)]{boquien09} Boquien, M., \etal\ 2009, \apj, 706, 553 
\bibitem[Boquien \etal(2012)]{boquien12} Boquien, M., \etal\ 2012, \aap, 539, A145 
\bibitem[Bouchet \etal(1985)]{bouchet85} Bouchet, P., Lequeux, J., Maurice, E., Prevot, L., \& Prevot-Burnichon, M.L. 1985, \aap, 149, 33
\bibitem[Bruzual(1983)]{bruzual83} Bruzual G. 1983, \apj, 273, 105
\bibitem[Buat \etal(2002)]{buat02} Buat, V., \etal\ 2002, \aap, 383, 801 
\bibitem[Buat \etal(2005)]{buat05} Buat, V., \etal\ 2005, \apj, 619, L51 
\bibitem[Buckalew \etal(2005)]{buckalew05} Buckalew, B.A., Kobulnicky, H.A., \& Dufour, R.J. 2005, \apjs, 157, 30
\bibitem[Burgarella \etal(2005)]{burgarella05} Burgarella, D., \etal\ 2005, \mnras, 360, 1413
\bibitem[Calzetti \etal(1994)]{calzetti94} Calzetti, D., Kinney, A.L., \& Storchi-Bergmann, T. 1994, \apj, 429, 582
\bibitem[Calzetti \etal(2000)]{calzetti00} Calzetti, D., \etal\ 2000, \apj, 533, 682
\bibitem[Calzetti(2001)]{calzetti01} Calzetti, D. 2001, \pasp, 113, 1449
\bibitem[Calzetti \etal(2005)]{calzetti05} Calzetti, D., \etal\ 2005, \apj, 633, 871
\bibitem[Calzetti \etal(2010)]{calzetti10} Calzetti, D., \etal\ 2010, \apj, 714, 1256
\bibitem[Christensen \etal(1997)]{christensen97} Christensen, T., Petersen, L. \& Gammelgaard, P. 1997, \aap, 322, 41
\bibitem[Cortese \etal(2006)]{cortese06} Cortese, L., \etal\ 2006, \apj, 637, 242
\bibitem[Cortese \etal(2008)]{cortese08} Cortese, L., \etal\ 2008, \mnras, 386, 1157
\bibitem[Corwin \etal(1994)]{corwin94} Corwin, H.G., Buta, R.J., \& de Vaucouleurs, G. 1994, \aj, 108, 2128
\bibitem[Croxall \etal(2009)]{croxall09} Croxall, K.V., \etal\ 2009, \apj, 705, 723
\bibitem[Dalcanton \etal(2009)]{dalcanton09} Dalcanton, J.J., \etal\ 2009, \apjs, 183, 67
\bibitem[Dale \etal(2002)]{dale02} Dale, D.A. \& Helou, G. 2002, \apj, 576, 159
\bibitem[Dale \etal(2007)]{dale07} Dale, D.A., \etal\ 2007, \apj, 655, 863 
\bibitem[Dale \etal(2009)]{dale09} Dale, D.A., \etal\ 2009, \apj, 703, 517 
\bibitem[de Vaucoulers \etal(1995)]{devaucouleurs95} de Vaucouleurs, G., de Vaucouleurs, A., Corwin, H.G., Buta, R.J., Paturel, G., \& Fouque, P. 1995, Third Reference Catalogue of Bright Galaxies (RC3), VizieR Online Data Catalog, 7155
\bibitem[Fazio \etal(2004)]{fazio04} Fazio, G.G., \etal\ 2004, \apjs, 154, 10 
\bibitem[Fitzpatrick(1999)]{fitzpatrick99} Fitzpatrick, E.L. 1999, PASP, 111, 63
\bibitem[Gallagher \& Hunter(1989)]{gallagher89} Gallagher, J.S. \& Hunter, D.A. 1989, \aj, 98, 806
\bibitem[Gil de Paz \etal(2000a)]{gildepaz00a} Gil de Paz, A., Zamorano, J., Gallego, J., \& de B. Dom\'inguez, F. 2000a, \aaps, 145, 377
\bibitem[Gil de Paz \etal(2000b)]{gildepaz00b} Gil de Paz, A., Zamorano, J., \& Gallego, J. 2000b, \aap, 361, 465
\bibitem[Gil de Paz \etal(2007)]{gildepaz07} Gil de Paz, A., \etal\ 2007, \apj, 661, 115
\bibitem[Gordon \etal(2000)]{gordon00} Gordon, K.D., \etal\ 2000, \apj, 533, 236
\bibitem[Gordon \etal(2004)]{gordon04} Gordon, K.D., \etal\ 2004, \apjs, 154, 215
\bibitem[Guseva \etal(2000)]{guseva00} Guseva, N.G., Izotov, Y.I., \& Thuan, T.X. 2000, \apj, 531, 776
\bibitem[Hao \etal(2011)]{hao11} Hao, C.-N., \etal\ 2011, \apj, 741, 124
\bibitem[Hidalgo-G\'amez \etal(2001)]{hidalgo01} Hidalgo-G\'amez, A.M., Masegosa, J., \& Olofsson, K. 2001, \aap, 369, 797
\bibitem[Hidalgo-G\'amez \& Olofsson(2002)]{hidalgo02} Hidalgo-G\'amez, A.M. \& Olofsson, K. 2002, \aap, 389, 836
\bibitem[Hopp \& Schulte-Ladbeck(1991)]{hopp91} Hopp, U. \& Schulte-Ladbeck, R.E. 1991, \aap, 248, 1
\bibitem[Hunter \etal(1982)]{hunter82} Hunter, D.A., Gallagher, J.S., \& Rautenkranz, D. 1982, \apjs, 49, 53
\bibitem[Hunter \& Gallagher(1985)]{hunter85} Hunter, D.A. \& Gallagher, J.S. 1985, \apjs, 58, 533
\bibitem[Hunter \& Hoffman(1999)]{hunter99} Hunter, D.A., \& Hoffman, L. 1999, \aj, 117, 2789
\bibitem[Inoue \etal(2006)]{inoue06} Inoue, A.K., \etal\ 2006, \mnras, 370, 380
\bibitem[Izotov \etal(2006)]{izotov06} Izotov, Y. , Stasi\'nska, G., Meynet, G., Guseva,  G., \& Thuan, T.X. 2006, \aap, 448, 955
\bibitem[Izotov \etal(2007)]{izotov07} Izotov, Y.I., Thuan, T.X., \& Stasi\'nska, G. 2007, \apj, 662, 15
\bibitem[Jester \etal(2005)]{jester05} Jester, S., \etal\ 2005, \aj, 130, 873
\bibitem[Johnson \etal(2007)]{johnson07} Johnson, B.D., \etal\ 2007, \apjs, 173, 377 
\bibitem[Kauffmann \etal(2003)]{kauffmann03} Kauffmann, G., \etal\ 2003, \mnras, 341, 33
\bibitem[Kennicutt \etal(1994)]{kennicutt94} Kennicutt, R.C., Tamblyn, P., Congdon, C.E. 1994, \apj, 435, 22
\bibitem[Kennicutt \& Skillman(2001)]{kennicutt01} Kennicutt, R.C., Jr. \& Skillman, E.D. 2001, \aj, 121, 1461
\bibitem[Kennicutt \etal(2008)]{kennicutt08} Kennicutt, R.C., Lee, J.C., Funes, J.G., Sakai, S., \& Akiyama, J.S. 2008, \apjs, 178, 247
\bibitem[Kewley \etal(2005)]{kewley05} Kewley, L.J., Jansen, R.A., \& Geller, M.J. 2005, \pasp, 117, 227
\bibitem[Kinman \& Davidson(1981)]{kinman81} Kinman, T.D. \& Davidson, K. 1981, \apj, 243, 127
\bibitem[Kniazev \etal(2003)]{kniazev03} Kniazev, A.Y., Grebel, E.K., Hao, L., Strauss, M.A., Brinkmann, J., \& Fukugita, M. 2003, \apj, 593, L73
\bibitem[Kniazev \etal(2004)]{kniazev04} Kniazev, A.Y., Pustilnik, S.A., Grebel, E.K., Lee, H., \& Pramskij, A.G. 2004, \apjs, 153, 429
\bibitem[Kobulnicky(1999)]{kibulnicky99} Kobulnicky, H.A., Kennicutt, R.C., \& Pizagno, J.L. 1999, \apj, 514, 544
\bibitem[Komatsu \etal(2011)]{komatsu11} Komatsu, E., \etal\ 2011, \apjs, 192, 18
\bibitem[Kong \etal(2004)]{kong04} Kong, X., Charlot, S., Brinchmann, J., Fall, S.M. 2004, \mnras, 349, 769
\bibitem[Lee \etal(2003)]{lee03} Lee, H., Grebel, E.K., \& Hodge, P.W. 2003, \aap, 401, 141
\bibitem[Lee \& Skillman(2004)]{lee04} Lee, H. \& Skillman, E.D. 2004, \apj, 614, 698
\bibitem[Lee \etal(2005)]{lee05} Lee, H., Skillman, E.D., \& Venn, K.A. 2005, \apj, 620, 223
\bibitem[Lee \etal(2009a)]{lee09a} Lee, J.C., \etal\ 2009a, \apj, 706, 599 
\bibitem[Lee \etal(2009b)]{lee09b} Lee, J.C., \etal\ 2009b, \apj, 692, 1305 
\bibitem[Lee \etal(2011)]{lee11} Lee, J.C., \etal\ 2011, \apjs, 192, 6 
\bibitem[Leitherer \etal(1999)]{leitherer99} Leitherer, C., \etal\ 1999, \apjs, 123, 3
\bibitem[Magrini \etal(2007)]{magrini07} Magrini, L., V\'ilchez, J.M., Mampaso, A., Corradi, R.L.M., \& Leisy, P. 2007, \aap, 470, 865
\bibitem[Mao \etal(2012)]{mao12} Mao, Y-.W., \etal\ 2012, \apj, 757, 52 
\bibitem[Martin(1997)]{martin97} Martin, C.L. 1997, \apj, 491, 561
\bibitem[Masegosa \etal(1994)]{masegosa94} Masegosa, J., Moles, M., \& Campos-Aguilar, A. 1994, \apj, 420, 576
\bibitem[McCall \etal(1985)]{mccall85} McCall, M.L.,  Rybski, P.M., \& Shields, G.A. 1985, \apjs, 57, 1
\bibitem[Meurer \etal(1999)]{meurer99} Meurer, G.R., Heckman, T.M., \& Calzetti, D. 1999, \apj, 521, 64
\bibitem[Miller \& Hodge(1996)]{millerh96} Miller, B.W. \& Hodge, P. 1996, \apj, 458, 467
\bibitem[Miller(1996)]{miller96} Miller, B.W. 1996, \aj, 112, 991
\bibitem[Moles \etal(1990)]{moles90} Moles, M., Aparicio, A., \& Masegosa, J. 1990, \aap, 228, 310
\bibitem[Mu\~{n}oz-mateos \etal(2009)]{munoz09} Mu\~{n}oz-mateos, J.C., \etal\ 2009, \apj, 701, 1965
\bibitem[Moustakas \& Kennicutt(2006)]{moustakas06} Moustakas, J., \& Kennicutt, R.C. 2006, \apj, 651, 155 
\bibitem[Moustakas \etal(2010)]{moustakas10} Moustakas, J. \etal, 2010, \apjs, 190, 233 
\bibitem[Overzier \etal(2011)]{overzier11} Overzier, R.A., \etal\ 2011, \apj, 726, L7
\bibitem[Panuzzo \etal(2007)]{panuzzo07} Panuzzo, P., \etal\ 2007, \mnras, 375, 640
\bibitem[P\'erez-Gonz\'alez \etal(2006)]{perez06} P\'erez-Gonz\'alez, P.G., \etal 2006, \apj, 648, 987
\bibitem[P\'erez-Montero \& D\'iaz(2003)]{perezmontero03} P\'erez-Montero, E. \& D\'iaz, A. 2003, \mnras, 346, 105
\bibitem[Popescu \etal(2005)]{popescu05} Popescu, C.C., \etal\ 2005, \apj, 619, L75
\bibitem[Pustilnik \etal(2003)]{pustilnik03} Pustilnik, S., Zasov, A., Kniazev, A., Pramskij, A., Ugryumov, A., \& Burenkov, A. 2003, \aap, 400, 841
\bibitem[Pustilnik \etal(2005)]{pustilnik05} Pustilnik, S.A., Kniazev, A.Y., \& Pramskij, A.G. 2005, \aap, 443, 91
\bibitem[Raimann \etal(2000)]{raimann00} Raimann, D., Storchi-Bergmann, T., Bica, E., Melnick, J., \& Schmitt, H. 2000, \mnras, 316, 559
\bibitem[Reddy \etal(2010)]{reddy10} Reddy, N.A., \etal\ 2010, \apj, 712, 1070
\bibitem[Reddy \etal(2012)]{reddy12} Reddy, N.A., \etal\ 2012, \apj, 744, 154
\bibitem[Rieke \etal(2004)]{rieke04} Rieke, G.H., \etal\ 2004, \apjs, 154, 25
\bibitem[Rosolowsky \& Simon(2008)]{rosolowsky08} Rosolowsky, E. \& Simon, J.D. 2008, \apj, 675, 1213
\bibitem[Saviane \etal(2008)]{saviane08} Saviane, I., Ivanov, V.D., Held, E.V., Alloin, D., Rich, R.M., Bresolin, F., \& Rizzi, L. 2008, \aap, 487, 901
\bibitem[Schlegel \etal(1998)]{schlegel98} Schlegel, D.J., Finkbeiner, D.P., \& Davis, M. 1998, \apj, 500, 525
\bibitem[Seibert \etal(2005)]{seibert05} Seibert, M., \etal\ 2005, \apj, 619, L55  
\bibitem[Skillman \etal(1989)]{skillman89} Skillman, E.D., Kennicutt, R.C., \& Hodge, P.W. 1989, \apj, 347, 875
\bibitem[Skillman \etal(1997)]{skillman97} Skillman, E.D., Bomans, D.J., \& Kobulnicky, H.A. 1997, \apj, 474, 205
\bibitem[Skillman \etal(2003)]{skillman03} Skillman, E.D., C\^ot\'e, S., \& Miller, B.W. 2003, \aj, 125, 610
\bibitem[Skrutskie \etal(2006)]{skutskie06} Skrutskie, M.F., Cutri, R.M., \& Heckman, T.M., \etal\ 2005, \apj, 619, L55
\bibitem[Stasi\'nska \etal(1986)]{stasinska86} Stasi\'nska, G., Comte, G., \& Vigroux, L. 1986, \aap, 154, 352
\bibitem[Storchi-Bergmann \etal(1994)]{storchi94}  Storchi-Bergmann, T., Calzetti, D., \& Kinney, A.L. 1994, \apj, 429, 572
\bibitem[Takeuchi \etal(2012)]{takeuchi12} Takeuchi, T.T., \etal\ 2012, \apj, 755, 144 
\bibitem[Taylor \etal(2005)]{taylor05} Taylor, V.A., Jansen, R.A., Windhorst, R.A.,  Odewahn, S.C., \& Hibbard, J.E. 2005, \apj, 630, 784
\bibitem[Thilker \etal(2007)]{thilker07} Thilker, D.A., \etal\ 2007, \apjs, 173, 572
\bibitem[Thuan \& Izotov(2005)]{thuan05} Thuan, T.X. \& Izotov, Y.I. 2005, \apjs, 161, 240
\bibitem[Tremonti \etal(2004)]{tremonti04} Tremonti, C.A., \etal\ 2004, \apj, 613, 898
\bibitem[T\"ullmann \etal(2003)]{tullmann03} T\"ullmann, R., Rosa, M.R., Elwert, T., Bomans, D.J., Ferguson, A.M.N., \& Dettmar, R.J. 2003, \aap, 412, 69 
\bibitem[van Zee \etal(1997a)]{vanzee97a} van Zee, L., Haynes, M.P., \& Salzer, J.J. 1997a, \aj, 114, 2479 
\bibitem[van Zee \etal(1997b)]{vanzee97b} van Zee, L., Haynes, M.P., \& Salzer, J.J. 1997b, \aj, 114, 2497 
\bibitem[van Zee \& Haynes(2006)]{vanzeeh06} van Zee, L. \& Haynes, M.P. 2006, \apj, 636, 214
\bibitem[van Zee \etal(2006)]{vanzee06} van Zee, L., Skillman, E.D., \& Haynes, M.P. 2006, \apj, 637, 269
\bibitem[Vila-Costas \& Edmunds(1993)]{vilacostas93} Vila-Costas, M.B. \& Edmunds, M.G. 1993, \mnras, 265, 199
\bibitem[Walsh \etal(1997)]{walsh97} Walsh, J.R. \& Roy, J.-R. 1997, \mnras, 288, 726
\bibitem[Whitmore \etal(1999)]{whitmore99} Whitmore, B.C., Zhang, Q., Leitherer, C., Fall, S.M., Schweizer, F., \& Miller, B.W. 1999, AJ, 118, 1551
\bibitem[Zaritsky \etal(1994)]{zaritsky94} Zaritsky, D., Kennicutt, R.C. Jr., \& Huchra, J.P. 1994, \apj, 420, 87
\bibitem[Zaritsky \etal(2004)]{zaritsky04} Zaritsky, D., Harris, J., Thompson, I. B., \& Grebel, E. K. 2004, AJ, 128, 1606
}
\end{thebibliography}
\end{document}